\documentclass[a4paper,fleqn,usenatbib]{mnras}

\usepackage{newtxtext,newtxmath}

\usepackage[T1]{fontenc}
\usepackage{ae,aecompl}

\usepackage{anyfontsize}
\usepackage{amsmath}
\usepackage{amsfonts}
\usepackage{amsbsy}

\usepackage{newtxtext}
\usepackage{newtxmath}

\usepackage{tensor}
\usepackage{mathrsfs}
\usepackage{bm}

\usepackage[utf8]{inputenc}

\usepackage{graphicx}
\usepackage{epsfig}
\usepackage{epstopdf}

\usepackage[usenames,dvipsnames]{xcolor}

\newcommand{\be}{\begin{equation}}
\newcommand{\ee}{\end{equation}}

\def\be{\begin{equation}}
\def\ee{\end{equation}}
\def\beq{\begin{eqnarray}}
\def\eeq{\end{eqnarray}}

\usepackage{graphicx}	
\usepackage{amsmath}	

\title[The role of plasma in tests of gravity]{The tune of the universe:\\ the role of plasma in tests of strong-field gravity}

\author[V. Cardoso, W-D Guo, C. F. B. Macedo and P. Pani]{
Vitor Cardoso,$^{1}$\thanks{E-mail: vitor.cardoso@ist.utl.pt}
Wen-Di Guo$^{2,1}$\thanks{E-mail: wen-di.guo@tecnico.ulisboa.pt}
Caio F. B. Macedo,$^{3}$\thanks{E-mail: caiomacedo@ufpa.br}
and Paolo Pani,$^{4}$\thanks{E-mail: paolo.pani@uniroma1.it}
\\
$^{1}$CENTRA, Departamento de Física, Instituto Superior Técnico – IST,Universidade de Lisboa – UL, Avenida Rovisco Pais 1, 1049 Lisboa, Portugal\\
$^{2}$Institute of Theoretical Physics $\&$ Research Center of Gravitation, Lanzhou University, Lanzhou 730000, China\\
$^{3}$Faculdade de Física, Universidade Federal do Par\'a, Salin\'opolis, Par\'a, 68721-000 Brazil\\
$^{4}$Dipartimento di Fisica, ``Sapienza'' Universit\`a di Roma \& Sezione INFN Roma1, Piazzale Aldo Moro 5, 00185, Roma, Italy
}

\begin{document}
\label{firstpage}
\pagerange{\pageref{firstpage}--\pageref{lastpage}}
\maketitle

\begin{abstract}
Gravitational-wave astronomy, together with precise pulsar timing and long baseline interferometry,
is changing our ability to perform tests of fundamental physics with astrophysical observations.
Some of these tests are based on electromagnetic probes or electrically charged bodies, and assume an empty universe.
However, the cosmos is filled with plasma, a dilute medium which prevents the propagation of
low-frequency, small-amplitude electromagnetic waves.
We show that the plasma hinders our ability to perform some strong-field gravity tests, in particular: (i)~nonlinear 
plasma effects dramatically quench plasma-driven superradiant instabilities; (ii)~the contribution of electromagnetic 
emission to the inspiral of charged black hole binaries is strongly suppressed; (iii)~electromagnetic-driven
secondary modes, although present in the spectrum of charged black holes, are excited to negligible amplitude in the 
gravitational-wave ringdown signal. The last two effects are relevant also in the case of massive fields that propagate 
in vacuum and can jeopardize tests of modified theories of gravity containing massive degrees of freedom.
\end{abstract}

\section{Introduction}
The birth of gravitational-wave~(GW) astronomy~\citep{LIGOScientific:2018mvr} together with optical/infrared
interferometry and radio very large baseline interferometry~\citep{2018A&A...618L..10G,Akiyama:2019cqa} opened the door
to new tests of General Relativity in the strong-field regime. Of special relevance are tests of the black-hole~(BH)
paradigm, including all the nontrivial general relativistic effects~\citep{Yagi:2016jml,Barack:2018yly,Cardoso:2019rvt}.
For example,
strong evidence for the existence of photon spheres was provided already by LIGO/Virgo~\citep{Cardoso:2019rvt} and by
the Event Horizon Telescope~\citep{Akiyama:2019cqa}. In parallel, strong-field gravity may offer routes to tests
of the dark matter content of our universe, turning compact objects into astrophysical particle detectors~\citep{Brito:2014wla,Baumann:2019ztm,Brito:2015oca}.

Several of these tests are directly or indirectly based on electromagnetic~(EM) probes or electrically charged objects,
and often assume that photons propagate freely in the universe. However, our universe is filled with plasma, a dilute
medium which prevents the propagation of low-frequency (and small-amplitude) EM waves.
The scope of this work is to revisit some strong-gravity phenomena to include the crucial effect played by the photon
coupling to plasma, and to discuss novel effects that have been so far neglected.
We consider constraints on the EM charge of compact objects and plasma-driven superradiant
instabilities~\citep{Pani:2013hpa,Brito:2015oca,Conlon:2017hhi}, briefly reviewed below.
Throughout this manuscript we use geometric units with $G=c=1$.
\subsection{Constraints on the charge of compact objects}
Given that EM fields are ubiquitous and play a key role in most of the known universe, it is only natural
to ask whether BHs or other compact objects are endowed with electric charge. Astrophysical BHs are considered to be
electrically neutral due to a variety of effects, including electron-positron production and neutralization by the
surrounding
plasma~\citep{Gibbons:1975kk,1969ApJ...157..869G,1975ApJ...196...51R,Blandford:1977ds,Barausse:2014tra,Barausse:2014pra}
. However, mergers
occur in violent conditions --~possibly including strong magnetic fields~-- and sufficiently far away that one may
question whether all conditions for neutrality are met~\citep{Cardoso:2016olt,Zajacek:2018vsj}. In addition, certain
dark matter candidates are (weakly) electrically charged and can circumvent the conditions for
neutralization~\citep{Cardoso:2016olt}.

Motivated by the potential of GW astronomy to explore these issues, the coalescence of charged BH {\it in electrovacuum}
was studied in recent years, both
nonlinearly~\citep{Zilhao:2012gp,Zilhao:2013nda,Liebling:2016orx,Jai-akson:2017ldo,Bozzola:2019aaw,Bozzola:2020mjx}
and perturbatively in the extreme mass ratio limit~\citep{Zhu:2018tzi}.

Neglecting environmental effects~\citep{Barausse:2014tra,Barausse:2014pra}, it was shown that GW observations of the
inspiral stage~\citep{Cardoso:2016olt,Christiansen:2020pnv} have the
potential to provide interesting constraints on the charge of BHs~\citep{Cardoso:2016olt,Liu:2020cds,Christiansen:2020pnv}.
Likewise, the ringdown stage can in principle be used to study
the EM charge of the remnant: gravitational and EM modes couple, and the characteristic GW vibration modes of the
system contain also a new, EM-like, family of modes, similar to the case of certain modified theories of gravity with
extra scalar degrees of freedom nonminimally coupled to
gravity~\citep{Molina:2010fb,Blazquez-Salcedo:2016enn,Okounkova:2017yby,Witek:2018dmd,Okounkova:2020rqw}.
Thus, it has been argued that the detection of two or more modes can in principle provide constraints on the
mass, spin, and charge of the final BH~\citep{Cardoso:2016olt}.

\subsection{Superradiance and the search for new fundamental fields}
Another very concrete example of the discovery potential of BHs and GW astronomy concerns new, fundamental ultralight
degrees of freedom. These would render spinning BHs  unstable, and lead to a transfer of
rotational energy to large-scale condensates in their surroundings; these time-varying structures would then emit
quasi-monochromatic GWs, a smoking-gun for new physics~\citep{Arvanitaki:2009fg,Arvanitaki:2010sy}
(see~\citep{Brito:2015oca} for a review).
The mechanism at work in superradiant instabilities requires two key ingredients: an ergoregion that ``forces'' the
field to be dragged along with the compact object, transfering energy and angular momentum to the
field~\citep{zeldovich1,zeldovich2,Brito:2015oca}, and a massive bosonic field. The field mass effectively
confines the entire setup, therefore turning an energy-extraction mechanism into an instability
mechanism~\citep{Damour:1976kh,Detweiler:1980uk,Brito:2015oca}. For a BH of mass $M$ and angular momentum $J=\chi M^2$, 
and a vector field of
mass $\hbar\mu$, such process may be very
efficient~\citep{Pani:2012vp,Pani:2012bp,Witek:2012tr,Endlich:2016jgc,East:2017mrj,East:2017ovw,Baryakhtar:2017ngi,
East:2018glu,Frolov:2018ezx,Dolan:2018dqv,Siemonsen:2019ebd,Baumann:2019eav}; the timescale for BH
spin-down --~and for the build-up of a massive vector condensate~-- is controlled by the parameter
\be
\gamma=M\mu\,,
\ee
and, in the $\gamma \ll 1$ regime, is of the order~\citep{Detweiler:1980uk,Cardoso:2005vk,Brito:2015oca}
\be
\tau\sim\frac{M}{\chi\gamma^{7}}\approx \frac{16}{\chi}\left(\frac{M}{1\,M_\odot}\right)
\left(\frac{0.01}{\gamma}\right)^{7} \,{\rm yr} \,.\label{tau_vector}
\ee
The instability is suppressed when the object spins down to $\chi\sim 2\gamma$.
Stringent constraints on the existence of new particles can then be imposed via (the lack of) observations of GWs
emitted by the bosonic condensate that develops around the BH, ``gaps''
in the mass-spin plane of BHs~\citep{Arvanitaki:2009fg,Arvanitaki:2010sy}, polarimetry
measurements~\citep{Plascencia:2017kca}, etc, although nonlinear photon effects such as pair production or 
couplings to Standard Model fields can 
reduce these bounds~\citep{Fukuda:2019ewf,Ikeda:2018nhb,Brito:2015oca}. A complete review of the status of the field is 
discussed in Ref.~\citep{Brito:2015oca}.

Surprisingly, the existence of interstellar plasma permeating the universe could provide an outstanding opportunity to test
the existence of ergoregions while simultaneously predicting, or explaining, new phenomena. The interaction between EM
waves and ions in a plasma changes the dispersion relation and the effective equations of motion of low-frequency
photons~\citep{Dendy,Kulsrud:1991jt}. The dispersion relation of the photon acquires an effective-mass term given by the 
plasma frequency~\citep{Hora,Kulsrud:1991jt,Pani:2013hpa},
\be
\omega_p=\sqrt{\frac{e^2 n_e}{\epsilon_0 m_e}}=1.8\times 10^3\left(\frac{n_e}{10^{-3}\,{\rm 
cm}^{-3}}\right)^{1/2}\,{\rm rad\,s^{-1}}\,,\label{typical_plasma_frequency}
\ee
where $n_e$ is the electron number density in the plasma, whereas $m_e$ and $e$ are the electron mass and charge,
respectively, and $\epsilon_0$ is the vacuum permeability.
It is clear from this that the effective mass would be interesting from the point of view of superradiant instabilities
of astrophysical systems, since the controlling parameter $\gamma$ is appreciable~\citep{Conlon:2017hhi},
\be
\gamma= \omega_p M\approx 0.09 \left(\frac{\omega_p}{1.8\times 10^3\,{\rm Hz}}\right)\left(\frac{M}{10M_\odot}\right)\,,\label{plasma_freq_dimensionless}
\ee
and therefore the associated instability timescale~\eqref{tau_vector} is relatively small.

Superradiant instabilities in the presence of plasmas have therefore been argued to produce important
signatures, including distortions in the cosmic background radiation from primordial BHs~\citep{Pani:2013hpa}, and have
been suggested as a possible explaination for Fast Radio Bursts from stellar-size BHs~\citep{Conlon:2017hhi}.
This has motivated further work on the topic, including the impact of a nonhomogeneous plasma
profile around a spinning BH, which reduces the instability rate~\citep{Dima:2020rzg}.
It has been also argued that the spin of neutron stars could be limited via the same plasma-driven
mechanism~\citep{Cardoso:2017kgn}, a truly tantalizing prospect to explain the systematic reduction of the spin measured 
in pulsars compared to the mass-shedding limit.

\subsection{The role of plasma: shortcomings of previous analyses}
Here, we argue that the aforementioned analyses have severe shortcomings, all related to the fact that they
neglected some key ingredient in the photon-plasma interaction.

Constraints on the EM charge have so far neglected the fact that plasma is not
transparent to low-frequency waves. The balance arguments used to understand how inspiral proceeds breakdown if EM
radiation is not being transported to null infinity. Note, in particular, that the plasma reflects any sufficiently 
low-frequency radiation. For the plasma frequency \eqref{typical_plasma_frequency}, this corresponds to reflecting back 
all radiation whose wavelength is {\it larger} than the binary separation. Thus, the usual boundary conditions involved 
in the retarded Green's function change and the nature of the solution is also completely different. This properties of 
a radiator inside a cavity have been studied from a quantum and classical 
perspective~\citep{HAROCHE1985347,Haroche_physics_today,DOWLING1991415} and support the above view; the radiation can be 
extremely suppressed or enhanced depending on the cavity and radiator~\footnote{\label{footnote1}We note that the relative size between 
the cavity and the radiation wavelength is important in the outcome; the boundary conditions are paramount: for 
absorbing cavities which are much larger than the wavelength of radiation, the backreaction on the radiator is expected 
to be negligible; this also is the physical reason why an absorption of photons from, say, the Sun, is not going 
to affect its emission properties.}. 

Furthermore, previous studies on GW spectroscopy 
for charged BHs also overlooked environmental effects
and how they impact the ringdown stage, in particular they neglected whether the plasma can affect the amplitude of 
EM-driven quasinormal modes.

Finally, the mechanisms leading to plasma-driven superradiant instabilities neglect
backreaction and nonlinear effects. In particular, a small growing electric field will re-arrange the
plasma distribution, changing its ``effective mass.'' In other words, the plasma is assumed to be totally opaque to such
low-frequency radiation, but such assumption needs justification, especially for unstable processes in which the 
EM field initially grows exponentially.

Here, we wish to examine these questions more closely.

\section{Nonlinear effects make plasma transparent to radiation}
Previous studies on plasma-driven BH superradiant instabilities are based on the assumption of a fixed constant plasma
density (see Ref.~\citep{Dima:2020rzg} for an extension to constant, non-homogenous density profiles with a scalar toy 
model) and, most notably, of weak electric
fields. However, relativistic and nonlinear effects cause waves of frequency
\be
\omega_p\left(1+\frac{e^2E^2}{m_e^2\omega^2}\right)^{-1/2}<\omega<\omega_p\,, \label{crit}
\ee
to propagate~\citep{1970PhFl...13..472K,1971PhRvL..27.1342M}. Here $E$ is the amplitude of the electric field of the
wave. Therefore, when the electric field is weak no frequency $\omega<\omega_p$ can propagate in the plasma, in
agreement with the linear analysis. However, there is a critical electric field
\be
E_{\rm crit}=\frac{m_e}{e}\sqrt{\omega_p^2 - \omega^2}\,, \label{Ecrit}
\ee
above which waves with frequency $\omega$ propagate into the plasma.

When dealing with superradiant instabilities this is a particularly important effect, because the electric field grows exponentially in the initial phase of the instability, before saturating
due to nonlinear effects. Indeed, since the dominant superradiant mode has roughly $\omega\sim
\omega_p\left(1-\frac{1}{2}\gamma^2\right)$~\citep{Pani:2012vp,Pani:2012bp,Pani:2013hpa}, in the $\gamma\ll1$ limit the
critical electric field that makes the plasma transparent to this mode is
\begin{equation}
E^{\rm SR}_{\rm crit}\sim \frac{m_e}{e}\omega_p \gamma =0.27\left(\frac{n_e}{10^{-3}{\rm 
cm}^{-3}}\right)^{1/2}\frac{\gamma}{0.09}\,{\rm V/m}\,. \label{EcritSR}
\end{equation}

Even in the absence of such nonlinear mechanism (but we do not have any reason to speculate on its absence), the 
assumption that low-frequency EM waves do not propagate in the plasma breaks
down when the plasma becomes hot and relativistic, i.e., when the collisional velocity $\sim 1$. Since the change in the
momentum of one electron over a time $\Delta t$ is $\Delta P=F\Delta t$, we get the critical electric field for this to
happen $E^{\rm rel}=m_e/(e\Delta \tau)$
with $\Delta \tau$ the mean collision time between electrons and ions in the plasma. For relativistic electrons, $\Delta \tau=\ell_e$,
with $\ell_e=n_e^{-1/3}$ being the mean separation. Therefore, neglecting nonlinear effects the critical value of the 
electric field above which plasma confinement breaks down reads
\be
E^{\rm SR}_{\rm rel}=\frac{m_en_e^{1/3}}{e}=5\times 10^6 \left(\frac{n_e}{10^{-3} {\rm cm}^{-3}}\right)^{1/3}\,{\rm V 
}\,{\rm m}^{-1}\,. \label{EcritREL}
\ee

When the electric field grows to the above values, the plasma becomes transparent, a burst ensues, and the
process starts anew\footnote{We note that the electric field of this radiation burst decreases with distance; for 
large enough distances the plasma will again be opaque to it; however, this is a subdominant effect in this 
process, and its overall consequence is to provide the plasma with a small --~and irrelevant for our 
discussion~-- effective mass.}.

One can estimate how much energy is removed from the rotating object in each cycle. The density of energy is $\sim
\epsilon_0 E^2$, and extended over a spatial distance $L_{\rm cloud}\sim 5M/\gamma^2$, set by the size 
of the bosonic cloud~\citep{Brito:2015oca}. Thus, the total energy in the condensate is
\be
U=\epsilon_0\left(E^{\rm SR}\right)^2L_{\rm cloud}^3\,,
\ee
for $E^{\rm SR}=E^{\rm SR}_{\rm crit}$ or $E^{\rm SR}=E^{\rm SR}_{\rm rel}$.
On the other hand, the rotational energy of a compact object of radius $R$ is $K\sim MR^2\Omega^2$,
with $R\sim 2M$ and $\Omega \gtrsim \gamma/M$. Using Eq.~\eqref{EcritSR}, one finds
\beq
\frac{U}{K}&\lesssim& \frac{125 M^2m_en_e}{4 \gamma^6} \\
&\approx&9\times 10^{-39}\left(\frac{M}{10M_{\odot}}\right)^2\left(\frac{n_e}{10^{-3} {\rm 
cm}^{-3}}\right)\left(\frac{0.09}{\gamma}\right)^6\,.\nonumber
\eeq

We note that, even in the absence of the above bound, the threshold (Eq.~\eqref{EcritREL}) at which the plasma becomes transparent under the assumption of relativistic electrons (thus emitting bremsstrahlung radiation and breaking all the assumptions leading to the plasma cutoff 
frequency) yields 
\beq
\frac{U}{K}&\lesssim& \frac{125\epsilon_0 M^2m_e^2n_e^{2/3}}{4\gamma^8e^2} \\
&\approx&3\times 10^{-24}\left(\frac{M}{10 M_{\odot}}\right)^2\left(\frac{n_e}{10^{-3} {\rm 
cm}^{-3}}\right)^{2/3}\left(\frac{0.09}{\gamma}\right)^8\,.\nonumber 
\eeq 

Therefore, the above simple analysis shows that relativistic and nonlinear effects hamper dramatically the tapping of
rotational energy from a spinning BH. In practice, immediately after the instability occurs the electric field in the
BH surroundings becomes large enough as to render the plasma transparent to low-frequency photons, making the trapping
(and, in turn, the whole instability) inefficient.

\section{Plasmas and merging charged BHs}
As previously mentioned, several mechanisms contribute to the neutralization of BHs. For example, charged BHs are 
quickly discharged by 
Hawking radiation or by
pair-production. In addition, a small amount of external plasma with total mass $M_{\rm plasma}\sim 10^{-18} M$ is 
sufficient to discharge a BH on a short timescale $\tau\sim 10^{-11}\,{\rm yr}$~\citep{Cardoso:2016olt}. Neutralization 
from surrounding plasma would normally screen charge on a Debye lengthscale: the charge of a BH is effectively screened
on the spatial scale (exceptions to this rule may occur with stars, but not with objects without 
atmospheres~\citep{1978ApJ...220..743B})
\be
\lambda_D=\sqrt{\frac{\epsilon_0 k_B T}{n_e e^2}}\approx 2 \left(\frac{T}{10^6\,K}\right)^{1/2}\left(\frac{10^{-3} {\rm 
cm}^{-3}}{n_e}\right)^{1/2} \,{\rm km}\,.
\ee
Colder or denser media yield even tighter atmospheres. Thus, for all practical purposes BHs surrounded by plasmas are neutral.

In the eventuality that the near-horizon region is plasma-depleted and that neutralization does not occur, then two BHs can indeed be
orbiting each other and emit EM radiation. However, also in this case the plasma surrounding the binary strongly 
hampers EM emission, as discussed in the next section.

\subsection{The inspiral phase}
The low-frequency, nearly Newtonian stage in the life of a binary of two compact objects
provides stringent constraints on the gravity theory and possible new interactions~\citep{Will:2014kxa,Yagi:2016jml}.
The underlying mechanism is as follows. Compact objects such as neutron stars or BHs on tight orbits
dissipate energy mostly through GW emission. Mass loss via winds, tidal acceleration,
or friction with the environment are all negligible in comparison. In these circumstances, the energy loss
can be computed using numerical methods or a post-Newtonian expansion. It is found that GW emission quickly 
circularizes the orbit~\citep{Peters:1964zz}.
It is thus customary to assume that the two orbiting objects are on quasi-circular orbits. To lowest 
order in the orbital velocity, the energy loss is given by the quadrupole formula
\be
\frac{d{\cal E}^{\rm grav}_{\rm rad}}{dt}=-\frac{32}{5}\frac{m_1^2m_2^2(m_1+m_2)}{L^5}\,,\label{rad_quad}
\ee
where $L$ is the orbital radius and $m_i$ is the mass of the $i$-th body. On the other hand, energy loss to radiation 
triggers an evolution of the binary 
parameters. In particular, this can be obtained using a quasi-adiabatic approximation to evolve the otherwise constants 
of motion such as the binary's energy and angular momentum. Such evolution can be determined using the expression for 
the orbital energy
\be
{\cal E}_{\rm orb}=-\frac{m_1 m_2}{2L}\,,\label{orb_relation}
\ee
equating $d{\cal E}_{\rm orb}/dt$ to $d{\cal E}_{\rm rad}/dt$ and promoting the orbital radius $L$ to be a function of time.

New physics will impact both the emission of radiation, changing Eq.~\eqref{rad_quad}, and the orbital relation 
\eqref{orb_relation}.
For example, if the objects carry electric charge there is emission
of mostly dipolar EM waves in addition to (mostly quadrupolar) GW emission. When the charge-to-mass ratio is small, 
dipolar radiation is of the order~\citep{Cardoso:2016olt,Liu:2020cds,Christiansen:2020pnv}
\be
\frac{d{\cal E}^{\rm elec}_{\rm 
rad}}{dt}=-\frac{2}{3}\left(\frac{Q_1}{m_1}-\frac{Q_2}{m_2}\right)^2\frac{m_1^2m_2^2}{4\pi \epsilon_0 L^4}\,,\label{rad_dip}
\ee
where $Q_i$ is the charge of the $i$-th body.
Thus, EM emission dominates at separations larger than
\begin{equation}
 L_{\rm crit}=\frac{48 M_{\rm tot}}{5(\delta_1-\delta_2)^2}\,, \label{Lcrit}
\end{equation}
where $M_{\rm tot}=m_1+m_2$ and $\delta_i=Q_i/(\sqrt{4\pi\epsilon_0}m_i)<1$ is the charge-to-mass ratio of the $i$-th body. For binaries at separation $L\gtrsim L_{\rm crit}$ the EM emission
causes a distinct evolution of the orbital phase as time progresses. These effects have be used to impose stringent constraints on the charge of the inspiralling 
objects~\citep{Cardoso:2016olt,Christiansen:2020pnv,Bozzola:2020mjx}.

Unfortunately, the Larmor result~\eqref{rad_dip} is valid in electrovacuum, but not in the presence of a plasma. It can 
be immediately recognized that the plasma frequency~\eqref{typical_plasma_frequency} is much larger than the orbital 
frequency of astrophysical BHs or neutron stars, especially in the relevant regime when $L\gtrsim L_{\rm crit}$. Thus, the assumption that the generated waves are able to travel 
freely and contribute to energy loss is incorrect. In fact, the entire calculation underlying \eqref{rad_dip}, in 
particular the imposition of Sommerfeld conditions at infinity, is not justified.

Is it possible that, just like the superradiance clouds in the previous section, photons are still able to tunnel 
through via nonlinear effects? We now show that this is not possible. Take a pointlike mass $m_i$, carrying charge $Q_i$, in motion with acceleration $a$.
Assuming Sommerfeld conditions at infinity, the electric field at distance $r$ in the wave zone is of order~\citep{jackson_classical_1999}
\be
E=\frac{Q_i}{4\pi\epsilon_0}\frac{a}{r}\,.
\ee
When the motion is circular and dictated by the inverse-square law, then $a\sim \Omega^2 L$, with $\Omega=\sqrt{M_{\rm tot}/L^3}$ being the Keplerian frequency. For simplicity we assume that the binary components are weakly charged, so that the acceleration is mostly provided by gravity (the calculation generalizes trivially).
Thus, at small $r$ the electric field is large and possibly larger than the critical field for transparency 
which, to be conservative, can be assumed to be the largest value among those discussed in the previous section, 
i.e. Eq.~\eqref{Ecrit}.
Taking into account the typical plasma frequency~\eqref{typical_plasma_frequency}, one concludes that $\omega\ll\omega_p$
during the inspiral of astrophysical BHs, especially in the large-separation regime where the dipolar terms arising 
from EM emission could dominate (i.e., $L>L_{\rm crit}$). Thus, one finds the distance $r_{\rm crit}$ below which the electric 
field of the radiation is larger than $E_{\rm crit}$,
\be
{r_{\rm crit}}\approx \delta_i \delta_e \left(\frac{M_{\rm tot}}{L}\right)^2\frac{1}{\omega_p}\,,
\ee
where $\delta_e\approx 2\times 10^{21}$ is the electron charge-to-mass ratio and for simplicity we assumed equal mass ratio, i.e. $M\approx M_{\rm tot}/2$. 
By evaluating the above formula at $L\approx L_{\rm crit}$ (Eq.~\eqref{Lcrit}), and using Eq.~\eqref{typical_plasma_frequency}, we obtain
\begin{equation}
 \frac{r_{\rm crit}}{\lambda} \approx 10^{19} \delta_i(\delta_1-\delta_2)^7 \left(\frac{M_\odot}{M_{\rm tot}}\right)
\end{equation}
where $\lambda=2\pi/\Omega$ is the radiation wavelength at $L=L_{\rm crit}$. Note the strong dependence on the charge-to-mass ratios $\delta$'s and the fact thar $r_{\rm crit}=0$ if $\delta_1=\delta_2$, since in this case dipolar emission is suppressed. 
For supermassive BHs ($M_{\rm tot}\approx 10^6 M_\odot$), one finds $\frac{r_{\rm crit}}{\lambda}\lesssim 1$ whenever $\delta_i={\cal O}(0.01)$ or smaller.
%
%
In other words, in the radiation zone, the electric field is always sub-critical. This means that the plasma is 
\emph{not} transparent to this radiation. For stellar-mass BHs or for larger $\delta$'s, the ratio $r_{\rm crit}/\lambda$ can be larger by orders of magnitude. Nevertheless, notice that plasma is not only absorbing, but also actually reflecting radiation. As we remarked in footnote~\ref{footnote1}, such aspect is crucial in the dynamics of objects within cavities: within a corresponding number of cycles the radiation would have reflected off and interacted with the binary, affecting its dynamics.

Finally, the above calculation and argument may also apply, in principle, to interactions other than the EM one. Take for example some ``hidden'' vector $V^\mu$ field coupled to the Maxwell sector~\citep{Cardoso:2016olt}. In such a case, interstellar (charged) dark matter and standard interstellar plasma would both work to provide an effective mass to the propagating field $V^{\mu}$, possibly turning the interstellar plasma opaque to radiation, thereby suppressing emission in the dark sector.
However, the effective mass for the dark sector is dependent on the coupling and on the mass of the carriers. Therefore, the above constraints may be evaded in some beyond-Standard-Model scenarios.

To summarize, the use of EM dipolar losses [Eq.~\eqref{rad_dip}] to constrain the charge (EM or even possibly of some other ``dark'' interaction) of astrophysical objects is 
not justified. Careful understanding of the role of plasma is necessary, but not available at this point.
Indeed, the problem of two orbiting charged particles in a plasma shares some similarities with that of two particles 
in a perfectly reflecting box. In the latter case energy cannot be radiated to infinity and stationary solutions where 
the orbit does not shrink exist (see, e.g., Ref.~\citep{Dias:2012tq} for the case of a binary in asymptotically anti 
de-Sitter spacetime). However, the case of the plasma is much more complex since: i)~the 
reflection of the radiation is frequency dependent; and ii)~as discussed before plasma is opaque only at the linear 
level, in reality we expect both the radiation and the binary motion to affect the plasma dynamics and profile around 
the binary, potentially modifying the propagation/absorption of EM waves.

\subsection{The ringdown stage}

The existence of couplings between two fields introduces mixing of modes in the ringdown. In the context of collisions of 
two charged BHs, gravitational and EM perturbations are coupled to each other. This leads to mode mixing in the 
ringdown, which is described by gravitational-led and EM-led modes. The former (resp., latter) are those that 
correspond to the standard gravitational (resp., EM) modes of a Reissner-Nordstrom BH in the neutral limit ($Q\to0$). 
We illustrate this effect in Appendix~\ref{app:ringdown}, where we overview the collision of two electrically charged 
BHs within a perturbative  approach. The effect is more pronounced when the coupling between the two sectors is 
large, which in the case of electrically charged BHs is related to the product of their charges.

To take plasma physics into account demands a more careful analysis, as one should consider the perturbations induced by the matter surrounding 
each BH and possibly the binary as a whole. We are currently unable to deal with this problem in full generality, 
which would require the analysis of plasma physics in the curved spacetime near a BH. 
However, we are interested in understanding one particular and crucial
aspect of plasma physics, which is the introduction of an effective mass for the photon.
The simplest toy model to 
mimic this scenario should be a massive vector coupled to a charged BH. Unfortunately, there are no BH 
solutions in Einstein-Proca theory~\citep{Bekenstein:1971hc,Bekenstein:1972ky}, which makes even this toy model 
uninteresting. We can get a grasp into the qualitative aspects of the problem by investigating another
proxy toy model, in which two fields are coupled together with a large (possibly effective) mass term.

\subsubsection{Toy model}
We wish to study how does spacetime react when couplings to (possibly effective) massive fields are considered. Following the discussions presented in the previous section,
this is precisely what happens when plasma is present in the surroundings of BHs. Our prototype model, which presents 
the main desired features, is dynamical Chern-Simons theory with a self-interacting potential~\citep{Alexander:2005jz}. 
In this theory the axial gravitational perturbations couple to a (pseudo)scalar field. The similarities between the physics of plasmas and 
the dynamical Chern-Simons theory come from the fact that both theories contain a field coupled to the gravitational one (a Maxwell field in the plasma case, and a scalar field in Chern-Simons gravity, respectively) and both feature a trapping mechanism due to the masses of these fields. The scalar-field mass is analogous to the effective mass provided by the plasma frequency. The strength of the coupling in the dynamical Chern-Simons case is controlled by a theory parameter, while
in the plasma scenario is related to the BH charge.

The action of the dynamical Chern-Simons theory is
\begin{eqnarray}\label{action}
S&=&\kappa \int d^4 x\sqrt{-g}R+\frac{\alpha}{4}\int d^4 x\sqrt{-g}\vartheta~{}^*RR\nonumber\\
&-&\frac{\beta}{2}\int d^4 x\sqrt{-g}[g^{ab}\nabla_{a}\vartheta\nabla_{b}\vartheta+V(\vartheta)]+S_{\rm mat}\,,
\end{eqnarray}
where $\vartheta$ is the scalar field coupled to gravity, and
\begin{eqnarray}
{}^*RR=\frac{1}{2}R_{abcd}\epsilon^{baef}{R^{cd}}_{ef}
\end{eqnarray}
is the Pontryagin term. The equation of motion can be obtained by varying the action with respect to the metric
$g_{ab}$ and the scalar field $\vartheta$, namely:
\begin{eqnarray}
G_{ab}+\frac{\alpha}{\kappa}C_{ab}-\frac{1}{2\kappa}{T^{\vartheta}}_{ab}&=&\frac{1}{2\kappa}T_{ab}\,, \label{field1}\\
\Box\vartheta-\frac{dV}{d\vartheta}&=&-\frac{\alpha}{4\beta}{}^*RR\,,\label{field2}
\end{eqnarray}
where $G_{ab}=R_{ab}-\frac{1}{2}g_{ab}R$ is the Einstein tensor,
${T^{\vartheta}}_{ab}=\beta\left[\vartheta_{;a}\vartheta_{;b}-
\frac{1}{2}g_{ab}\Box\vartheta-g_{ab}V(\vartheta)\right]$, $\Box=g^{ab}\nabla_{a}\nabla_{b}$ is the D'Alembertian
operator, and 
\begin{eqnarray}
C^{ab}=\vartheta_{;c}\epsilon^{cde(a}\nabla_{e}{R^{b)}}_{d}+\vartheta_{;cd}{}^*R^{d(ab)c},
\end{eqnarray}
where $\vartheta_{;a}=\nabla_a\vartheta$, $\vartheta_{;ab}=\nabla_a\nabla_b\vartheta$, and
${}^*R^{abcd}=\frac{1}{2}\epsilon^{cdef}{R^{ab}}_{ef}$. In the geometric units adopted so far, $\kappa=1/16\pi$.

For concreteness, we consider the simplest potential for a massive scalar field, 
\begin{equation}
 V(\vartheta)=\frac{\mu^2}{2}\vartheta^2\,,
\end{equation}
where $\mu \hbar$ is the mass of the scalar.

We consider a spherically symmetric background, which is described by the Schwarzschild solution also in this theory.
Gravitational perturbations on this background can be expanded in a basis of (scalar, vector, and tensor) spherical
harmonics~\citep{Regge:1957td,Zerilli:1971wd,MTB}. The scalar field can be expanded in scalar spherical harmonics as follows
\begin{eqnarray}
\vartheta=\frac{\Theta^{lm}}{r}Y^{lm}e^{-i\omega t}.
\end{eqnarray}

While the polar (Zerilli) sector is the same as in General Relativity and reduced to a single, second-order, radial 
differential equation~\citep{Zerilli:1971wd,MTB}, the axial (Regge-Wheeler) sector is coupled to scalar 
perturbation and reduces to the system~\citep{Molina:2010fb} (we omit for simplicity the indices $l$ and $m$ in the 
variables $\Psi$ and $\Theta$),
%
\begin{eqnarray}
\frac{d^2}{dr_\star^2}\Psi+(\omega^2-V_{11})\Psi&=&
V_{12}\Theta\,,\label{eqq2}\\
\frac{d^2}{dr_\star^2}\Theta+(\omega^2-V_{22})\Theta&=&V_{21}\Psi,\label{eqpsi2}
\end{eqnarray}
where we have defined\footnote{It is worth noting that the couplings $\alpha$ and $\beta$ are degenerate and 
one of them can be set to unity without loss of generality~\citep{Molina:2010fb}. We use this freedom to fix $\alpha=1$ 
and keep $\beta$ as a free parameter.} 
\begin{eqnarray}
V_{11}&=&f(r)\left[\frac{l(l+1)}{r^2}-\frac{6M}{r^3}\right],\\
V_{12}&=&f(r)\frac{96\pi M}{r^5},\\
V_{21}&=&f(r)\frac{6M(l+2)!}{r^5\beta(l-2)!},\\
V_{22}&=&f(r)\left[\frac{l(l+1)}{r^2}\left(1+\frac{576\pi M^2}{\beta r^6}\right)+\frac{2M}{r^3}+\mu^2\right]\,,
\end{eqnarray}
with $f(r)=1-2M/r$, and $r_\star=r+2M\ln(r/2M -1)$ being the tortoise coordinate. 
In the above, $\Psi$ is the Fourier-transform of the Regge-Wheeler master function, itself a combination of (axial-like) metric fluctuations~\citep{Zerilli:1971wd,MTB,Molina:2010fb}.

From the above equations, one immediately sees that the scalar perturbations source the gravitational ones with a 
relative coupling $\propto \Theta/\Psi$, whereas the gravitational perturbations source the scalar ones with a relative 
coupling  $\propto \Psi/(\beta \Theta)$. Therefore, when $\beta$ is small (large), the scalar field is strongly 
(weakly) sourced by the gravitational perturbation, while the latter depends on $\beta$ only 
indirectly through the value of $\Theta$.

When using such Chern-Simons theory to understand plasmas around Schwarzschild BHs, one needs to control the 
couplings $M\mu$ and $\tilde \beta\equiv \beta M^4$.
The mass coupling parameter $M\mu$ depends on the environmental plasma, and can be estimated from Eq.~\eqref{plasma_freq_dimensionless}.
The coupling $\beta$ in Eqs.~\eqref{eqq2}-\eqref{eqpsi2} can be estimated by considering how charged perturbations 
couple to gravity. This coupling is insensitive to the plasma and should be well described by the Einstein-Maxwell 
theory, which we understand well. The relevant perturbation equations can be separated and decoupled, and reduced to the 
form of Eqs.~\eqref{eqq2}-\eqref{eqpsi2}~\citep{Cardoso:2016olt}. We find a coupling $\tilde\beta \sim 
M^2/Q^2\geq1$. Thus, $\tilde \beta=100,\,1$, correspond respectively to $Q=0.1,1\,M$. These values are 
representative of the constraints that could in principle be achieved with GW astronomy, as reported in previous work~\citep{Cardoso:2016olt}.

\subsubsection{Numerical procedure}

We have studied Eqs.~\eqref{eqq2}-\eqref{eqpsi2} with two different methods. In particular, we used established techniques to look
for the characteristic modes (quasinormal modes) of such a system in the 
frequency domain~\citep{Berti:2009kk,Pani:2013pma}. These frequencies tell us how fluctuations decay at very late times.
However, such a knowledge is insufficient, if not accompanied by the relative amplitudes of such fields. To understand this aspect, we also performed evolutions
of the corresponding 1+1 partial differential equations: following Refs.~\citep{Molina:2010fb,Macedo:2018txb}, we use the light-cone coordinates, $u=t-r_\star$ and 
$v=t+r_\star$. Then, the field equations can be written in matricial form as $4 \frac{\partial ^2}{\partial u \partial 
v} \Phi=-V \Phi$, with
\begin{eqnarray}
\Phi=\left(\begin{array}{c}\Psi\\ \Theta\end{array}\right),~~~~~ V=\left(\begin{array}{cc}V_{11}&V_{12}\\V_{21}&V_{22}
\end{array}\right).
\end{eqnarray}

We consider three different initial data ${\rm ID}_{2,\Psi,\Theta}$ of the form
\begin{eqnarray}
\Phi(0,v)=\left(\begin{array}{c}\epsilon_{\Psi}e^{-(v-v_c)^2/2\sigma}\\\epsilon_{\Theta}e^{-(v-v_c)^2/2\sigma}
\end{array}\right),~~~~~ \Phi(u,0)=\left(\begin{array}{c}0 \\0 \end{array}\right)\,, \nonumber
\end{eqnarray}
with $(\epsilon_{\Psi},\epsilon_{\Theta})=(1,1), \, (1,0),\, (0,1)$ for ${\rm ID}_{2},{\rm ID}_{\Psi},{\rm
ID}_{\Theta}$, respectively. In other words, ${\rm ID}_{2}$ corresponds to both fields initially perturbed, whereas 
${\rm ID}_{\Psi}$ and ${\rm ID}_{\Theta}$ correspond to only $\Psi$ or only $\Theta$ initially perturbed, respectively.
We focus on width $\sigma/M=1$ and a Gaussian located at $v_c/M=10$. The ranges of $u$ and $v$ are both $(0,400M)$,
and we extract the data at $r_\star=50M$. 
\subsubsection{Results}
%
\begin{figure*}[ht]
\includegraphics[width=\linewidth]{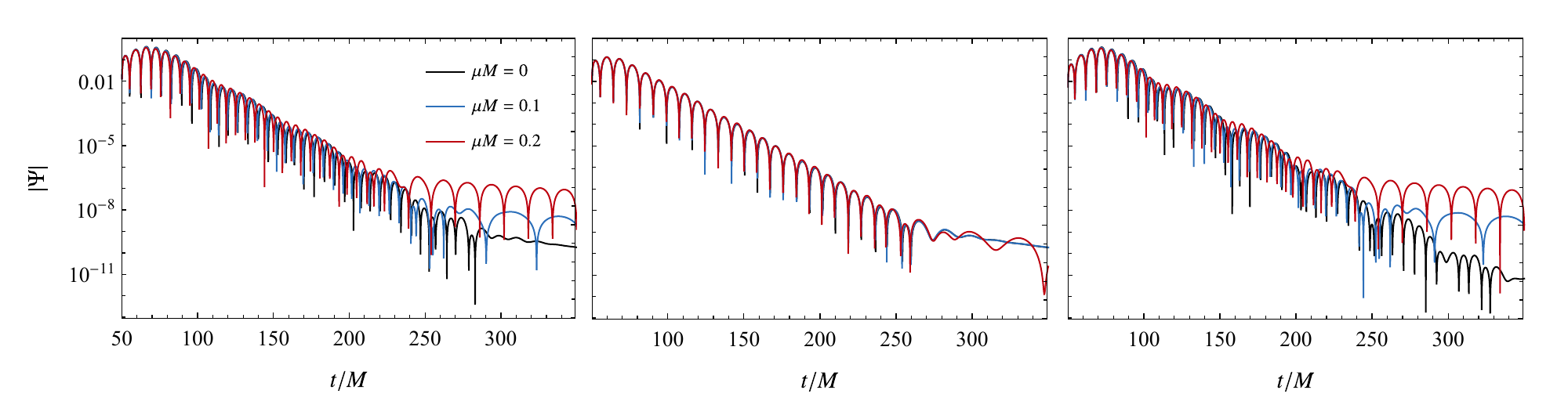}\\
\includegraphics[width=\linewidth]{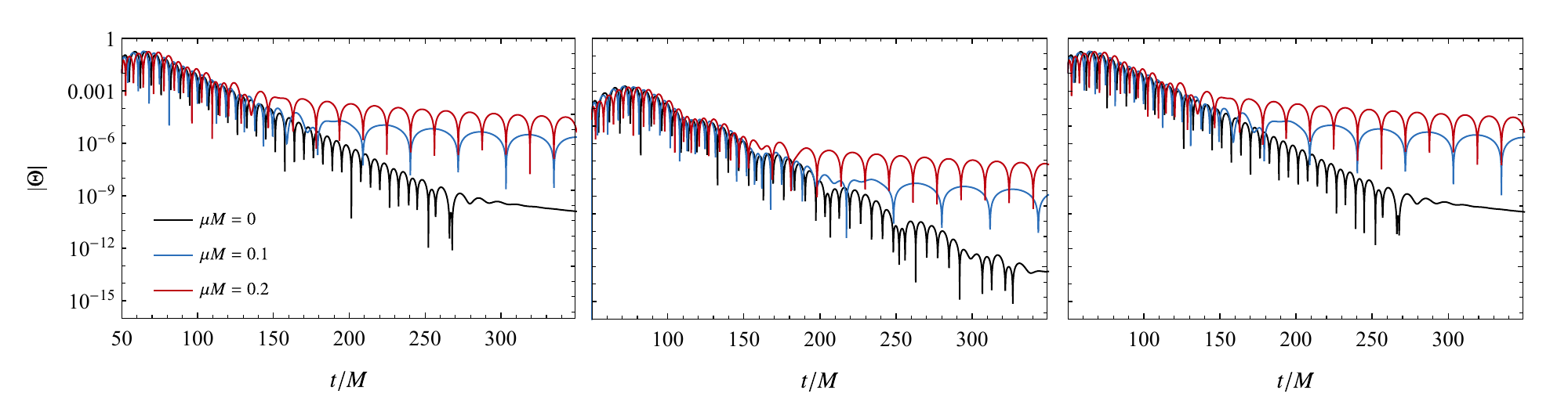}\\
\includegraphics[width=\linewidth]{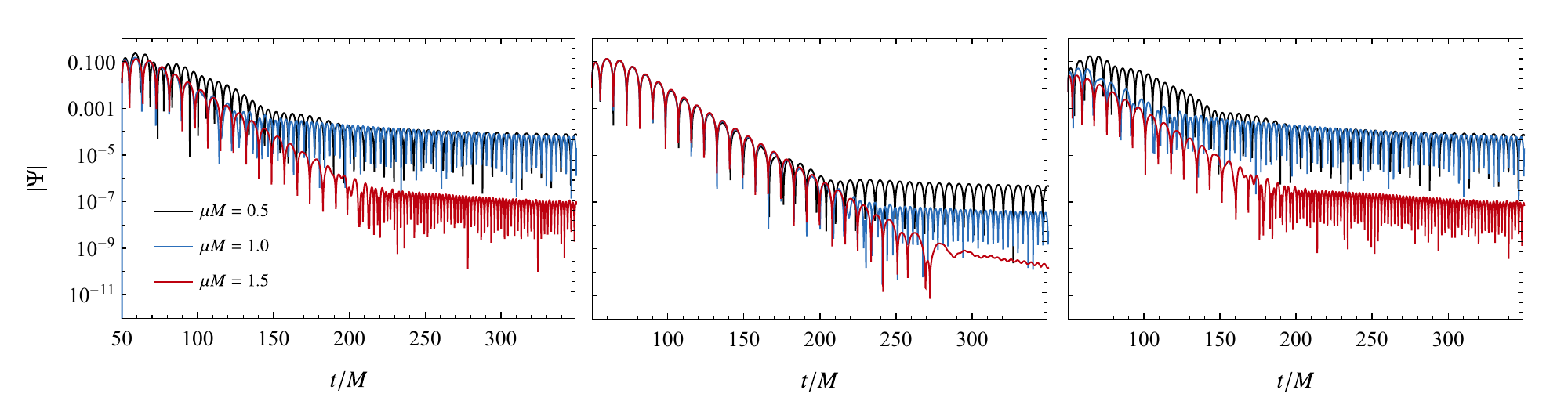}\\
\includegraphics[width=\linewidth]{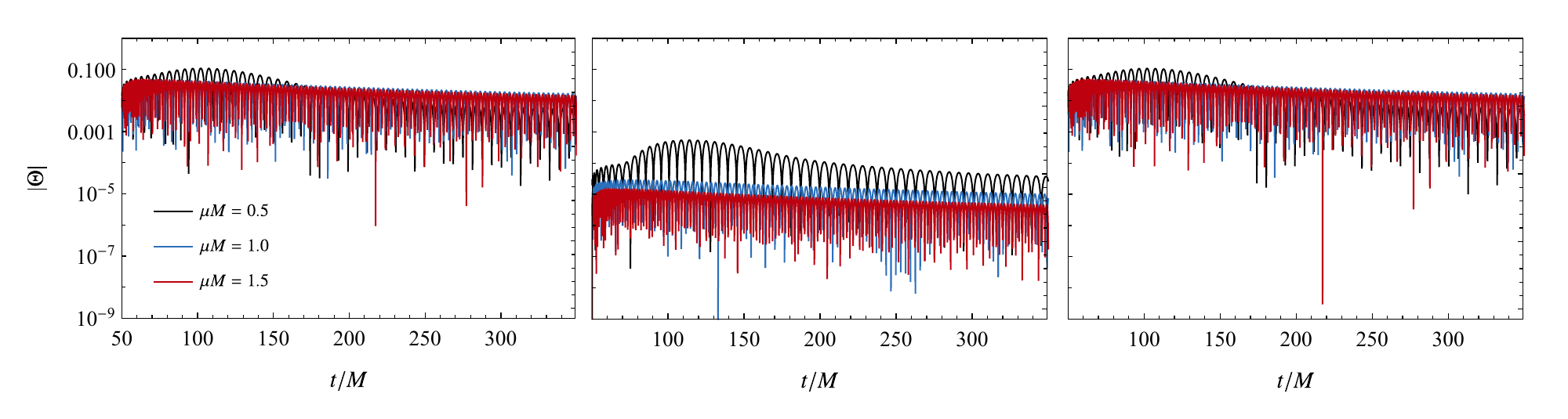}
\caption{Ringdown time evolution in our toy model in which a 
massive scalar field $\Theta$ is coupled to axial gravitational perturbations $\Psi$. We consider $\tilde \beta=100$ 
(small 
coupling) and the initial data is located at $v_c=10$. The top two row refer to $\mu M=0,0.1,0.2$ (zero or small mass 
term), whereas the bottom two rows correspond to $\mu M=0.5, 1, 1.5$ (moderately large mass term). From left to right: 
${\rm ID}_{2}$, ${\rm ID}_{\Psi}$, ${\rm ID}_{\Theta}$. In all cases we set $\alpha=1$.
\label{fig:beta100}}
\end{figure*}
For $\tilde \beta=100$, a frequency-domain analysis of the quadrupole modes with $l=2$ and mass coupling $M\mu=0.1$ 
indicates the presence of various characteristic modes. One is located at 
$M\omega\sim 0.369-0.092i$ and can be identified with a gravitational-led mode. In the decoupled $\tilde \beta \to 
\infty$ 
limit, this is the lowest and dominant gravitational quasinormal mode~\citep{Berti:2009kk,rdweb}. A second dominant mode is 
located at $M\omega\sim 0.506-0.094i$ which in the massless decoupled limit is the lowest quasinormal frequency of a 
minimally coupled scalar field. For $\tilde \beta=1$ the modes' frequencies change. The lowest gravitational-driven mode 
is at $0.29-0.098i$, while the scalar-driven goes up substantially to $M\omega \sim 1.4-0.14i$.

A mass term gives rise to new late-time phenomena, in particular the appearance of quasi-bound, Hydrogen-like states~\citep{Detweiler:1980uk,Brito:2015oca,Macedo:2018txb}.
These are followed by late-time power-law tails, triggered by back-scattering off spacetime curvature~\citep{Price:1971fb,Hod:1998ra,Koyama:2001ee,Koyama:2001qw,Witek:2012tr}.
For massless fields, the late-time power law tail is of the form $\Psi \sim t^{-(2l+3)}$ for scalars, for 
example~\citep{Price:1971fb,Witek:2012tr}. For massive fields, these tails have the form $t^p\sin\mu t$, with 
$p=-(l+3/2),\,-5/6$ at intermediate and late times, respectively. The quasi-bound state and tail phase can be seen to 
take over after $t\sim 250 M$ in the top panels of Fig.~\ref{fig:beta100}. Note that the very late-time behavior of 
massless fields is markedly different, as we remarked it is a pure power-law decay. We will not consider this stage in 
great detail, since we are mostly interested in the initial ringdown stage. For $\tilde \beta=100$ and $\mu=0.1$ we 
find $M\omega\sim 0.0999-2\times 10^{-17}i$. In general this mode is characterized by ${\rm Re}(\omega)\sim \mu$.
The quasi-bound state for $\tilde \beta=1$ remains at $\omega\sim \mu$, with the imaginary part 
increasing~\citep{Macedo:2018txb}.

\begin{figure*}[ht]
\includegraphics[width=\linewidth]{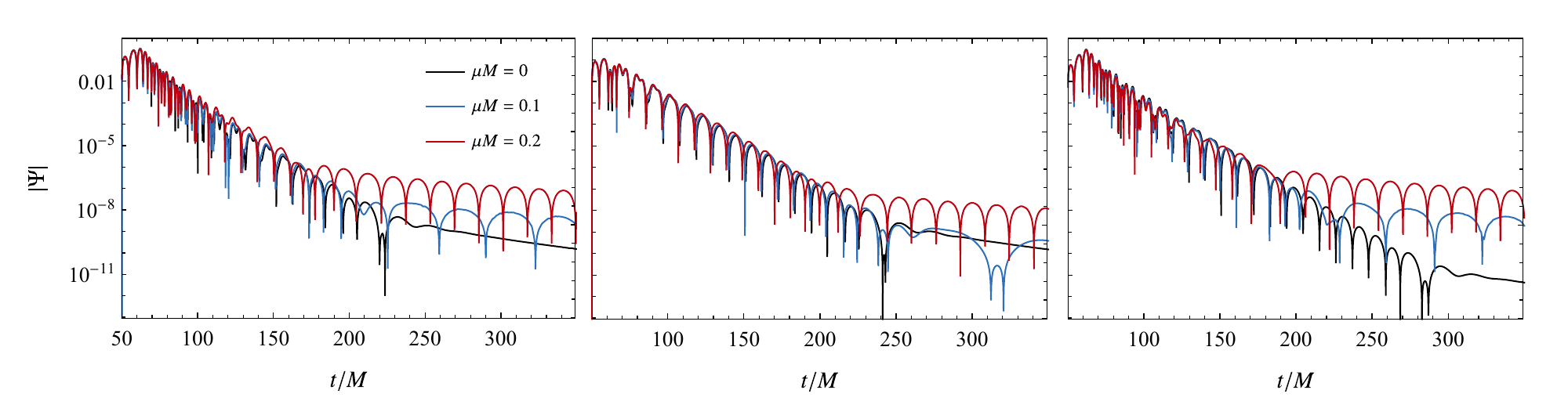}\\
\includegraphics[width=\linewidth]{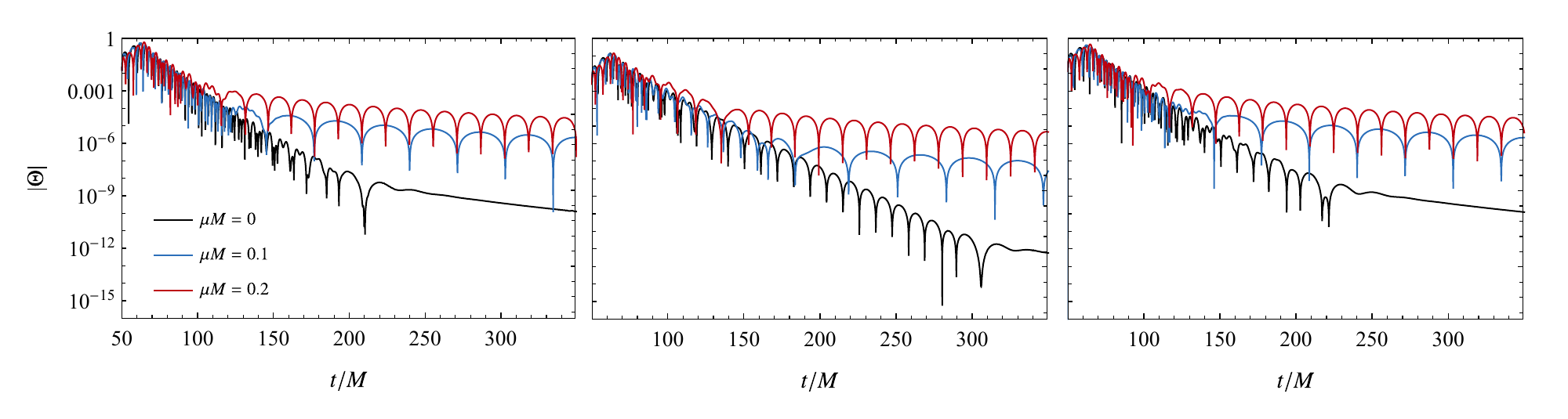}\\
\includegraphics[width=\linewidth]{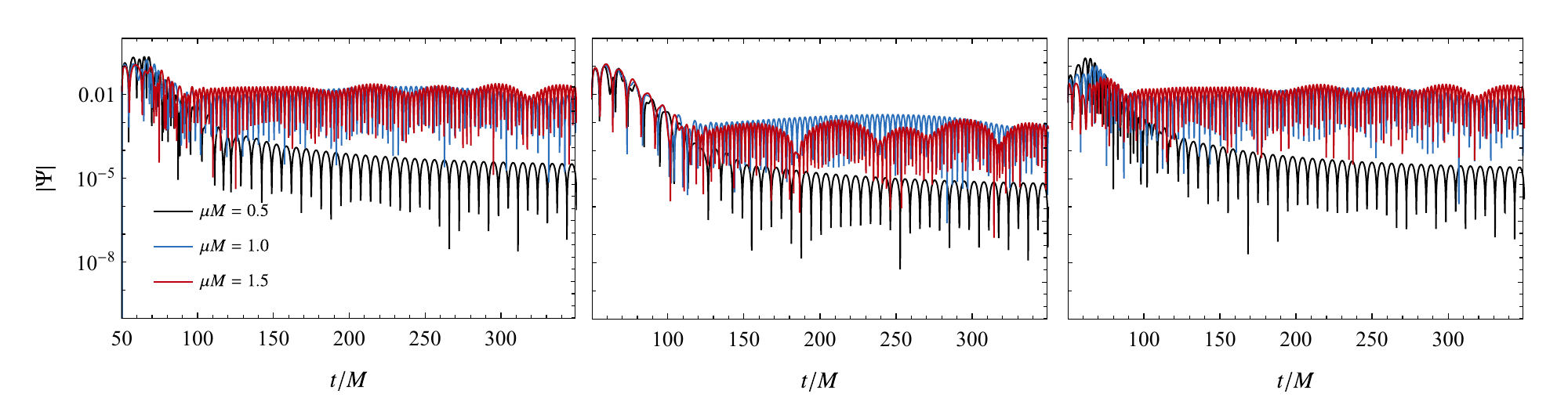}\\
\includegraphics[width=\linewidth]{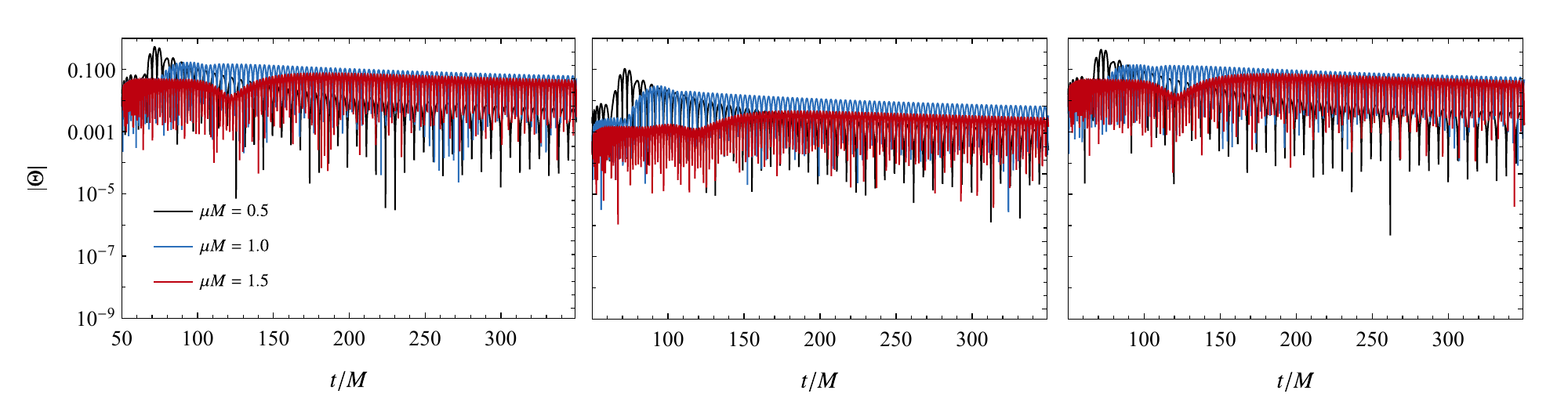}
\caption{Same as Fig.~\ref{fig:beta100} but for $\tilde \beta=1$ (stronger coupling). Notice that the high-frequency 
scalar-led mode
is present in the gravitational signal for low-mass couplings $M\mu$ but is absent when $M\mu$ is large.  
\label{fig:beta1}}
\end{figure*}

To which extent are these modes excited during the evolution of our initial conditions? The outcome of the time 
evolutions are summarized in Fig.~\ref{fig:beta100} (for $\tilde \beta=100$) and 
Fig.~\ref{fig:beta1} (for $\tilde \beta=1$) for different mass couplings $M\mu$. The left, middle, and right columns 
correspond to 
${\rm ID}_{2}$, ${\rm ID}_{\Psi}$, ${\rm ID}_{\Theta}$, respectively, whereas the top (bottom) two rows correspond to 
zero or small (large) values of the scalar mass coupling $\mu M$.

Although we show the evolution of both the scalar and the gravitational field, for the purpose of this discussion let us focus
on the gravitational sector alone, the first and third row in Figs.~\ref{fig:beta100} - \ref{fig:beta1}.
The first aspect that stands out is that, in general, the new channel --~the scalar~-- has an impact also in the 
gravitational sector.
The evolution of the gravitational field leads to a ringdown stage which is not a simple damped sinusoid, but a superposition
of at least two of the modes discussed above. Thus, the scalar quasinormal mode percolates to the gravitational sector due to the coupling. 
Furthermore, for massive scalar field, the very late time behavior of the gravitational sector is that of a weakly 
damped sinusoid controlled by the scalar quasi-bound states, ringing at
a frequency $\omega \sim \mu$~\citep{Macedo:2018txb}.

However, our results show that

\noindent {\bf (i)} The {\it amplitude} of the scalar-driven quasinormal mode in the gravitational sector is smaller at smaller couplings (large $\beta$).
This is also apparent from the panels in Figs.~\ref{fig:beta100} - \ref{fig:beta1}.

\noindent {\bf (ii)} More importantly, {\it at small couplings and large masses} the gravitational ringdown is universal. To a good approximation it corresponds to the modes of BHs in vacuum GR. This is depicted in the third row, second column of Fig.~\ref{fig:beta100}. Increasing the mass term $M\mu$ delays the appearance of the quasi-bound state dominance, where the fied rings at $\omega \sim \mu$, a clear imprint from the scalar sector in the gravitational waveform. Notice that even at large couplings this feature is present: the larger the mass $M\mu$ the more pure and scalar-free is the gravitational waveform. As in the rest of this work, this feature arises because the scalar is unable to propagate and therefore the equations decouple in practice.

\noindent {\bf (iii)} The above behavior holds well even when initially there is only a scalar field, such as in the 
third row, third column of Figs.~\ref{fig:beta100} and \ref{fig:beta1}.
The ``contamination'' of the gravitational wave by the scalar mode is smaller at large mass couplings. In other words, 
for large $M\mu$ the amplitude of the induced EM-led, gravitational mode
is small and decreases when $M\mu$ increases.

%

\section{Conclusions}

Recent years have witnessed new developments in strong-field
tests of gravity based on charged
binaries~\citep{Zilhao:2012gp,Zilhao:2013nda,Cardoso:2016olt,Liebling:2016orx,Jai-akson:2017ldo,Zhu:2018tzi,
Bozzola:2019aaw,Bozzola:2020mjx,Christiansen:2020pnv} and on plasma-driven superradiant
effects~\citep{Pani:2013hpa,Conlon:2017hhi,Dima:2020rzg}.
These tests either neglect the effect of plasma surrounding the binaries, or neglect nonlinear plasma-photon
interactions, or anyway treat the photon-plasma coupling in a simplistic way. Unfortunately, as shown in this work, EM 
emission from binary coalescence, secondary EM-driven modes in the ringdown, and plasma-driven instabilities are 
strongly suppressed when a more rigorous treatment is performed.

We argue that previous constraints and effects should be revised on the light of our results, 
and urge for a more detailed treatment of the photon-plasma interactions in strongly-gravitating systems, which is 
unavailable at the moment. Given the variety of scales and the complex physics involved, full numerical simulations 
using general-relativistic magneto-hydrodynamics might be required.

Although the main focus of this work was on plasma physics, some of our results are also relevant for tests of modified 
gravity~\citep{Berti:2015itd}. In particular, the suppression of the extra modes in the ringdown and of the dipolar 
emission in the inspiral should qualitatively apply also to the case of extra fundamental massive fields that propagate 
in vacuum. While it is well-known that a massive field suppresses the emission of low-frequency waves (see, e.g., 
Ref.~\citep{Cardoso:2011xi,Yunes:2011aa,Alsing:2011er,Cardoso:2013fwa,Cardoso:2013opa,Ramazanoglu:2016kul} for examples 
in scalar-tensor theory), we also predict that if the field is massive enough its modes --~while present in the spectrum 
of a BH remnant~-- cannot be sufficiently excited during the merger and are, therefore, undetectable. It is interesting 
that the relevant parameter for these effects is typically the coupling $\mu M$, which is huge for astrophysical BHs 
and the typical masses of particles in the standard model\footnote{For example, $\mu M\approx 
7\times 10^9$ for a particle as light as 
$\mu\hbar=1\,{\rm eV}$ and a stellar-mass BH.}. Therefore, if a putative extra fundamental field is massive 
(approximately with mass $\mu \hbar\gtrsim 10^{-10}\,{\rm eV}$ so that $\mu M\gtrsim 1$ for a stellar-mass BH) its 
effects in the inspiral and ringdown are strongly suppressed.

After submission of this work, the effects of plasma on the development of superradiant instabilities was also studied, from different viewpoints,
in Refs.~\citep{Blas:2020kaa,Blas:2020nbs}.

\section*{Acknowledgements}
%
We are indebted to Luis O. Silva and Jorge Vieira for discussions on the physics of plasmas, and for a number of useful references.
We thank Lang Liu and Rodrigo Vicente for useful feedback.
V. C. would like to thank Waseda University for warm hospitality and support
while this work was finalized.
V. C. acknowledges financial support provided under the European Union's H2020 ERC
Consolidator Grant ``Matter and strong-field gravity: New frontiers in Einstein's
theory'' grant agreement no. MaGRaTh--646597.
W. D. Guo acknowledges the financial support provided by the
scholarship granted by the Chinese Scholarship Council (CSC) and the Fundamental Research Funds
for the Central Universities (Grants Nos. lzujbky-2019-it21).
C.F.B.M acknowledges Conselho Nacional de Desenvolvimento Científico e Tecnológico (CNPq), and 
Coordenação de Aperfeiçoamento de Pessoal de Nível Superior (CAPES), from Brazil.
P.P. acknowledges financial support provided under the European Union's H2020 ERC, Starting 
Grant agreement no.~DarkGRA--757480, and under the MIUR PRIN and FARE programmes (GW-NEXT, CUP:~B84I20000100001), and 
support from the Amaldi Research Center funded by the MIUR program ``Dipartimento di Eccellenza'' (CUP: 
B81I18001170001).
This project has received funding from the European Union's Horizon 2020 research and innovation
programme under the Marie Sklodowska-Curie grant agreement No 690904.
We thank FCT for financial support through Project~No.~UIDB/00099/2020 and through grant PTDC/MAT-APL/30043/2017.
The authors would like to acknowledge networking support by the GWverse COST Action
CA16104, ``Black holes, gravitational waves and fundamental physics.''
%
%

\section*{Data availability}

The data underlying this article will be shared on reasonable request to the corresponding author.

\bibliographystyle{mnras}
\bibliography{References} 

\begin{thebibliography}{}
\makeatletter
\relax
\def\mn@urlcharsother{\let\do\@makeother \do\$\do\&\do\#\do\^\do\_\do\%\do\~}
\def\mn@doi{\begingroup\mn@urlcharsother \@ifnextchar [ {\mn@doi@}
  {\mn@doi@[]}}
\def\mn@doi@[#1]#2{\def\@tempa{#1}\ifx\@tempa\@empty \href
  {http://dx.doi.org/#2} {doi:#2}\else \href {http://dx.doi.org/#2} {#1}\fi
  \endgroup}
\def\mn@eprint#1#2{\mn@eprint@#1:#2::\@nil}
\def\mn@eprint@arXiv#1{\href {http://arxiv.org/abs/#1} {{\tt arXiv:#1}}}
\def\mn@eprint@dblp#1{\href {http://dblp.uni-trier.de/rec/bibtex/#1.xml}
  {dblp:#1}}
\def\mn@eprint@#1:#2:#3:#4\@nil{\def\@tempa {#1}\def\@tempb {#2}\def\@tempc
  {#3}\ifx \@tempc \@empty \let \@tempc \@tempb \let \@tempb \@tempa \fi \ifx
  \@tempb \@empty \def\@tempb {arXiv}\fi \@ifundefined
  {mn@eprint@\@tempb}{\@tempb:\@tempc}{\expandafter \expandafter \csname
  mn@eprint@\@tempb\endcsname \expandafter{\@tempc}}}

\bibitem[\protect\citeauthoryear{Abbott et~al.}{Abbott
  et~al.}{2019}]{LIGOScientific:2018mvr}
Abbott B.~P.,  et~al., 2019, \mn@doi [Phys. Rev.] {10.1103/PhysRevX.9.031040},
  X9, 031040

\bibitem[\protect\citeauthoryear{Abuter et~al.}{Abuter
  et~al.}{2018}]{2018A&A...618L..10G}
Abuter R.,  et~al., 2018, \mn@doi [\aap] {10.1051/0004-6361/201834294}, \href
  {https://ui.adsabs.harvard.edu/abs/2018A&A...618L..10G} {618, L10}

\bibitem[\protect\citeauthoryear{Akiyama et~al.}{Akiyama
  et~al.}{2019}]{Akiyama:2019cqa}
Akiyama K.,  et~al., 2019, \mn@doi [Astrophys. J.] {10.3847/2041-8213/ab0ec7},
  875, L1

\bibitem[\protect\citeauthoryear{Alexander}{Alexander}{2005}]{Alexander:2005jz}
Alexander T.,  2005, \mn@doi [Phys. Rept.] {10.1016/j.physrep.2005.08.002},
  419, 65

\bibitem[\protect\citeauthoryear{Alsing, Berti, Will  \& Zaglauer}{Alsing
  et~al.}{2012}]{Alsing:2011er}
Alsing J.,  Berti E.,  Will C.~M.,   Zaglauer H.,  2012, \mn@doi [Phys. Rev. D]
  {10.1103/PhysRevD.85.064041}, 85, 064041

\bibitem[\protect\citeauthoryear{Arvanitaki \& Dubovsky}{Arvanitaki \&
  Dubovsky}{2011}]{Arvanitaki:2010sy}
Arvanitaki A.,  Dubovsky S.,  2011, \mn@doi [Phys.Rev.]
  {10.1103/PhysRevD.83.044026}, D83, 044026

\bibitem[\protect\citeauthoryear{Arvanitaki, Dimopoulos, Dubovsky, Kaloper  \&
  March-Russell}{Arvanitaki et~al.}{2010}]{Arvanitaki:2009fg}
Arvanitaki A.,  Dimopoulos S.,  Dubovsky S.,  Kaloper N.,   March-Russell J.,
  2010, \mn@doi [Phys. Rev.] {10.1103/PhysRevD.81.123530}, D81, 123530

\bibitem[\protect\citeauthoryear{{Bally} \& {Harrison}}{{Bally} \&
  {Harrison}}{1978}]{1978ApJ...220..743B}
{Bally} J.,  {Harrison} E.~R.,  1978, \mn@doi [\apj] {10.1086/155961}, \href
  {https://ui.adsabs.harvard.edu/abs/1978ApJ...220..743B} {220, 743}

\bibitem[\protect\citeauthoryear{Barack et~al.}{Barack
  et~al.}{2019}]{Barack:2018yly}
Barack L.,  et~al., 2019, \mn@doi [Class. Quant. Grav.]
  {10.1088/1361-6382/ab0587}, 36, 143001

\bibitem[\protect\citeauthoryear{Barausse, Cardoso  \& Pani}{Barausse
  et~al.}{2014}]{Barausse:2014tra}
Barausse E.,  Cardoso V.,   Pani P.,  2014, \mn@doi [Phys. Rev.]
  {10.1103/PhysRevD.89.104059}, D89, 104059

\bibitem[\protect\citeauthoryear{Barausse, Cardoso  \& Pani}{Barausse
  et~al.}{2015}]{Barausse:2014pra}
Barausse E.,  Cardoso V.,   Pani P.,  2015, \mn@doi [J. Phys. Conf. Ser.]
  {10.1088/1742-6596/610/1/012044}, 610, 012044

\bibitem[\protect\citeauthoryear{Baryakhtar, Lasenby  \& Teo}{Baryakhtar
  et~al.}{2017}]{Baryakhtar:2017ngi}
Baryakhtar M.,  Lasenby R.,   Teo M.,  2017, \mn@doi [Phys. Rev.]
  {10.1103/PhysRevD.96.035019}, D96, 035019

\bibitem[\protect\citeauthoryear{Baumann, Chia, Stout  \& ter Haar}{Baumann
  et~al.}{2019}]{Baumann:2019eav}
Baumann D.,  Chia H.~S.,  Stout J.,   ter Haar L.,  2019, \mn@doi [JCAP]
  {10.1088/1475-7516/2019/12/006}, 12, 006

\bibitem[\protect\citeauthoryear{Baumann, Chia, Porto  \& Stout}{Baumann
  et~al.}{2020}]{Baumann:2019ztm}
Baumann D.,  Chia H.~S.,  Porto R.~A.,   Stout J.,  2020, \mn@doi [Phys. Rev.
  D] {10.1103/PhysRevD.101.083019}, 101, 083019

\bibitem[\protect\citeauthoryear{Bekenstein}{Bekenstein}{1972a}]{Bekenstein:1971hc}
Bekenstein J.~D.,  1972a, \mn@doi [Phys. Rev. D] {10.1103/PhysRevD.5.1239}, 5,
  1239

\bibitem[\protect\citeauthoryear{Bekenstein}{Bekenstein}{1972b}]{Bekenstein:1972ky}
Bekenstein J.,  1972b, \mn@doi [Phys. Rev. D] {10.1103/PhysRevD.5.2403}, 5,
  2403

\bibitem[\protect\citeauthoryear{Berti, Cardoso  \& Starinets}{Berti
  et~al.}{2009}]{Berti:2009kk}
Berti E.,  Cardoso V.,   Starinets A.~O.,  2009, \mn@doi [Class. Quantum Grav.]
  {10.1088/0264-9381/26/16/163001}, 26, 163001

\bibitem[\protect\citeauthoryear{Berti et~al.}{Berti
  et~al.}{2015}]{Berti:2015itd}
Berti E.,  et~al., 2015, \mn@doi [Class. Quant. Grav.]
  {10.1088/0264-9381/32/24/243001}, 32, 243001

\bibitem[\protect\citeauthoryear{Blandford \& Znajek}{Blandford \&
  Znajek}{1977}]{Blandford:1977ds}
Blandford R.,  Znajek R.,  1977, Mon.Not.Roy.Astron.Soc., 179, 433

\bibitem[\protect\citeauthoryear{Blas \& Witte}{Blas \&
  Witte}{2020a}]{Blas:2020kaa}
Blas D.,  Witte S.~J.,  2020a, \mn@doi [Phys. Rev. D]
  {10.1103/PhysRevD.102.123018}, 102, 123018

\bibitem[\protect\citeauthoryear{Blas \& Witte}{Blas \&
  Witte}{2020b}]{Blas:2020nbs}
Blas D.,  Witte S.~J.,  2020b, \mn@doi [Phys. Rev. D]
  {10.1103/PhysRevD.102.103018}, 102, 103018

\bibitem[\protect\citeauthoryear{Blazquez-Salcedo, Macedo, Cardoso, Ferrari,
  Gualtieri, Khoo, Kunz  \& Pani}{Blazquez-Salcedo
  et~al.}{2016}]{Blazquez-Salcedo:2016enn}
Blazquez-Salcedo J.~L.,  Macedo C. F.~B.,  Cardoso V.,  Ferrari V.,  Gualtieri
  L.,  Khoo F.~S.,  Kunz J.,   Pani P.,  2016, \mn@doi [Phys. Rev.]
  {10.1103/PhysRevD.94.104024}, D94, 104024

\bibitem[\protect\citeauthoryear{Bozzola \& Paschalidis}{Bozzola \&
  Paschalidis}{2019}]{Bozzola:2019aaw}
Bozzola G.,  Paschalidis V.,  2019, \mn@doi [Phys. Rev.]
  {10.1103/PhysRevD.99.104044}, D99, 104044

\bibitem[\protect\citeauthoryear{Bozzola \& Paschalidis}{Bozzola \&
  Paschalidis}{2021}]{Bozzola:2020mjx}
Bozzola G.,  Paschalidis V.,  2021, \mn@doi [Phys. Rev. Lett.]
  {10.1103/PhysRevLett.126.041103}, 126, 041103

\bibitem[\protect\citeauthoryear{Brito, Cardoso  \& Pani}{Brito
  et~al.}{2015}]{Brito:2014wla}
Brito R.,  Cardoso V.,   Pani P.,  2015, \mn@doi [Class. Quant. Grav.]
  {10.1088/0264-9381/32/13/134001}, 32, 134001

\bibitem[\protect\citeauthoryear{Brito, Cardoso  \& Pani}{Brito
  et~al.}{2020}]{Brito:2015oca}
Brito R.,  Cardoso V.,   Pani P.,  2020, {Superradiance}: {New Frontiers in
  Black Hole Physics}.
 Vol. 971, Springer (\mn@eprint {arXiv} {1501.06570}),
  \mn@doi{10.1007/978-3-319-19000-6}

\bibitem[\protect\citeauthoryear{Cardoso \& Pani}{Cardoso \&
  Pani}{2019}]{Cardoso:2019rvt}
Cardoso V.,  Pani P.,  2019, \mn@doi [Living Rev. Rel.]
  {10.1007/s41114-019-0020-4}, 22, 4

\bibitem[\protect\citeauthoryear{Cardoso \& Yoshida}{Cardoso \&
  Yoshida}{2005}]{Cardoso:2005vk}
Cardoso V.,  Yoshida S.,  2005, \mn@doi [JHEP] {10.1088/1126-6708/2005/07/009},
  07, 009

\bibitem[\protect\citeauthoryear{Cardoso, Lemos  \& Yoshida}{Cardoso
  et~al.}{2003}]{Cardoso:2003cn}
Cardoso V.,  Lemos J. P.~S.,   Yoshida S.,  2003, \mn@doi [Phys. Rev.]
  {10.1103/PhysRevD.68.084011}, D68, 084011

\bibitem[\protect\citeauthoryear{Cardoso, Chakrabarti, Pani, Berti  \&
  Gualtieri}{Cardoso et~al.}{2011}]{Cardoso:2011xi}
Cardoso V.,  Chakrabarti S.,  Pani P.,  Berti E.,   Gualtieri L.,  2011,
  \mn@doi [Phys. Rev. Lett.] {10.1103/PhysRevLett.107.241101}, 107, 241101

\bibitem[\protect\citeauthoryear{Cardoso, Carucci, Pani  \& Sotiriou}{Cardoso
  et~al.}{2013a}]{Cardoso:2013opa}
Cardoso V.,  Carucci I.~P.,  Pani P.,   Sotiriou T.~P.,  2013a, \mn@doi [Phys.
  Rev. D] {10.1103/PhysRevD.88.044056}, 88, 044056

\bibitem[\protect\citeauthoryear{Cardoso, Carucci, Pani  \& Sotiriou}{Cardoso
  et~al.}{2013b}]{Cardoso:2013fwa}
Cardoso V.,  Carucci I.~P.,  Pani P.,   Sotiriou T.~P.,  2013b, \mn@doi [Phys.
  Rev. Lett.] {10.1103/PhysRevLett.111.111101}, 111, 111101

\bibitem[\protect\citeauthoryear{Cardoso, Macedo, Pani  \& Ferrari}{Cardoso
  et~al.}{2016}]{Cardoso:2016olt}
Cardoso V.,  Macedo C. F.~B.,  Pani P.,   Ferrari V.,  2016, \mn@doi [JCAP]
  {10.1088/1475-7516/2016/05/054}, 1605, 054

\bibitem[\protect\citeauthoryear{Cardoso, Pani  \& Yu}{Cardoso
  et~al.}{2017}]{Cardoso:2017kgn}
Cardoso V.,  Pani P.,   Yu T.-T.,  2017, \mn@doi [Phys. Rev.]
  {10.1103/PhysRevD.95.124056}, D95, 124056

\bibitem[\protect\citeauthoryear{Chandrasekhar}{Chandrasekhar}{1983}]{MTB}
Chandrasekhar S.,  1983, The Mathematical Theory of Black Holes.
Oxford University Press, New York

\bibitem[\protect\citeauthoryear{Christiansen, Jim\'enez  \& Mota}{Christiansen
  et~al.}{2020}]{Christiansen:2020pnv}
Christiansen O.,  Jim\'enez J.~B.,   Mota D.~F.,  2020

\bibitem[\protect\citeauthoryear{Conlon \& Herdeiro}{Conlon \&
  Herdeiro}{2018}]{Conlon:2017hhi}
Conlon J.~P.,  Herdeiro C. A.~R.,  2018, \mn@doi [Phys. Lett.]
  {10.1016/j.physletb.2018.02.073}, B780, 169

\bibitem[\protect\citeauthoryear{Damour, Deruelle  \& Ruffini}{Damour
  et~al.}{1976}]{Damour:1976kh}
Damour T.,  Deruelle N.,   Ruffini R.,  1976, \mn@doi [Lett. Nuovo Cim.]
  {10.1007/BF02725534}, 15, 257

\bibitem[\protect\citeauthoryear{Dendy}{Dendy}{1989}]{Dendy}
Dendy R.~O.,  1989, {Plasma Dynamics}.
Oxford University Press, Oxford, UK

\bibitem[\protect\citeauthoryear{Detweiler}{Detweiler}{1980}]{Detweiler:1980uk}
Detweiler S.~L.,  1980, \mn@doi [Phys. Rev.] {10.1103/PhysRevD.22.2323}, D22,
  2323

\bibitem[\protect\citeauthoryear{Dias, Horowitz, Marolf  \& Santos}{Dias
  et~al.}{2012}]{Dias:2012tq}
Dias O.~J.,  Horowitz G.~T.,  Marolf D.,   Santos J.~E.,  2012, \mn@doi [Class.
  Quant. Grav.] {10.1088/0264-9381/29/23/235019}, 29, 235019

\bibitem[\protect\citeauthoryear{Dima \& Barausse}{Dima \&
  Barausse}{2020}]{Dima:2020rzg}
Dima A.,  Barausse E.,  2020, \mn@doi [Class. Quant. Grav.]
  {10.1088/1361-6382/ab9ce0}, 37, 175006

\bibitem[\protect\citeauthoryear{Dolan}{Dolan}{2018}]{Dolan:2018dqv}
Dolan S.~R.,  2018, \mn@doi [Phys. Rev.] {10.1103/PhysRevD.98.104006}, D98,
  104006

\bibitem[\protect\citeauthoryear{Dowling, Scully  \& DeMartini}{Dowling
  et~al.}{1991}]{DOWLING1991415}
Dowling J.~P.,  Scully M.~O.,   DeMartini F.,  1991, \mn@doi [Optics
  Communications] {https://doi.org/10.1016/0030-4018(91)90351-D}, 82, 415

\bibitem[\protect\citeauthoryear{East}{East}{2017}]{East:2017mrj}
East W.~E.,  2017, \mn@doi [Phys. Rev.] {10.1103/PhysRevD.96.024004}, D96,
  024004

\bibitem[\protect\citeauthoryear{East}{East}{2018}]{East:2018glu}
East W.~E.,  2018, \mn@doi [Phys. Rev. Lett.] {10.1103/PhysRevLett.121.131104},
  121, 131104

\bibitem[\protect\citeauthoryear{East \& Pretorius}{East \&
  Pretorius}{2017}]{East:2017ovw}
East W.~E.,  Pretorius F.,  2017, \mn@doi [Phys. Rev. Lett.]
  {10.1103/PhysRevLett.119.041101}, 119, 041101

\bibitem[\protect\citeauthoryear{Endlich \& Penco}{Endlich \&
  Penco}{2017}]{Endlich:2016jgc}
Endlich S.,  Penco R.,  2017, \mn@doi [JHEP] {10.1007/JHEP05(2017)052}, 05, 052

\bibitem[\protect\citeauthoryear{Frolov, Krtous, Kubiznak  \& Santos}{Frolov
  et~al.}{2018}]{Frolov:2018ezx}
Frolov V.~P.,  Krtous P.,  Kubiznak D.,   Santos J.~E.,  2018, \mn@doi [Phys.
  Rev. Lett.] {10.1103/PhysRevLett.120.231103}, 120, 231103

\bibitem[\protect\citeauthoryear{Fukuda \& Nakayama}{Fukuda \&
  Nakayama}{2020}]{Fukuda:2019ewf}
Fukuda H.,  Nakayama K.,  2020, \mn@doi [JHEP] {10.1007/JHEP01(2020)128}, 01,
  128

\bibitem[\protect\citeauthoryear{Gibbons}{Gibbons}{1975}]{Gibbons:1975kk}
Gibbons G.,  1975, \mn@doi [Commun.Math.Phys.] {10.1007/BF01609829}, 44, 245

\bibitem[\protect\citeauthoryear{{Goldreich} \& {Julian}}{{Goldreich} \&
  {Julian}}{1969}]{1969ApJ...157..869G}
{Goldreich} P.,  {Julian} W.~H.,  1969, \mn@doi [\apj] {10.1086/150119}, \href
  {http://adsabs.harvard.edu/abs/1969ApJ...157..869G} {157, 869}

\bibitem[\protect\citeauthoryear{Haroche \& Kleppner}{Haroche \&
  Kleppner}{1989}]{Haroche_physics_today}
Haroche S.,  Kleppner D.,  1989, \mn@doi [Physics Today]
  {https://doi.org/10.1063/1.881201}, 42, 24

\bibitem[\protect\citeauthoryear{Haroche \& Raimond}{Haroche \&
  Raimond}{1985}]{HAROCHE1985347}
Haroche S.,  Raimond J.,  1985, Academic Press, pp 347 -- 411,
  \mn@doi{https://doi.org/10.1016/S0065-2199(08)60271-7}, \url
  {http://www.sciencedirect.com/science/article/pii/S0065219908602717}

\bibitem[\protect\citeauthoryear{Hod \& Piran}{Hod \& Piran}{1998}]{Hod:1998ra}
Hod S.,  Piran T.,  1998, \mn@doi [Phys. Rev. D] {10.1103/PhysRevD.58.044018},
  58, 044018

\bibitem[\protect\citeauthoryear{Hora}{Hora}{1991}]{Hora}
Hora H.,  1991, {Plasmas at High Temperature and Density}.
Springer, Heidelberg

\bibitem[\protect\citeauthoryear{Ikeda, Brito  \& Cardoso}{Ikeda
  et~al.}{2019}]{Ikeda:2018nhb}
Ikeda T.,  Brito R.,   Cardoso V.,  2019, \mn@doi [Phys. Rev. Lett.]
  {10.1103/PhysRevLett.122.081101}, 122, 081101

\bibitem[\protect\citeauthoryear{Jackson}{Jackson}{1999}]{jackson_classical_1999}
Jackson J.~D.,  1999, Classical electrodynamics, 3rd ed. edn.
Wiley, New York, {NY}, \url {http://cdsweb.cern.ch/record/490457}

\bibitem[\protect\citeauthoryear{Jai-akson, Chatrabhuti, Evnin  \&
  Lehner}{Jai-akson et~al.}{2017}]{Jai-akson:2017ldo}
Jai-akson P.,  Chatrabhuti A.,  Evnin O.,   Lehner L.,  2017, \mn@doi [Phys.
  Rev.] {10.1103/PhysRevD.96.044031}, D96, 044031

\bibitem[\protect\citeauthoryear{Johnston, Ruffini  \& Zerilli}{Johnston
  et~al.}{1973}]{Johnston:1973cd}
Johnston M.,  Ruffini R.,   Zerilli F.,  1973, \mn@doi [Phys. Rev. Lett.]
  {10.1103/PhysRevLett.31.1317}, 31, 1317

\bibitem[\protect\citeauthoryear{{Kaw} \& {Dawson}}{{Kaw} \&
  {Dawson}}{1970}]{1970PhFl...13..472K}
{Kaw} P.,  {Dawson} J.,  1970, \mn@doi [Physics of Fluids] {10.1063/1.1692942},
  \href {http://adsabs.harvard.edu/abs/1970PhFl...13..472K} {13, 472}

\bibitem[\protect\citeauthoryear{Koyama \& Tomimatsu}{Koyama \&
  Tomimatsu}{2001}]{Koyama:2001ee}
Koyama H.,  Tomimatsu A.,  2001, \mn@doi [Phys. Rev. D]
  {10.1103/PhysRevD.64.044014}, 64, 044014

\bibitem[\protect\citeauthoryear{Koyama \& Tomimatsu}{Koyama \&
  Tomimatsu}{2002}]{Koyama:2001qw}
Koyama H.,  Tomimatsu A.,  2002, \mn@doi [Phys. Rev. D]
  {10.1103/PhysRevD.65.084031}, 65, 084031

\bibitem[\protect\citeauthoryear{Kulsrud \& Loeb}{Kulsrud \&
  Loeb}{1992}]{Kulsrud:1991jt}
Kulsrud R.,  Loeb A.,  1992, \mn@doi [Phys. Rev. D] {10.1103/PhysRevD.45.525},
  45, 525

\bibitem[\protect\citeauthoryear{Liebling \& Palenzuela}{Liebling \&
  Palenzuela}{2016}]{Liebling:2016orx}
Liebling S.~L.,  Palenzuela C.,  2016, \mn@doi [Phys. Rev.]
  {10.1103/PhysRevD.94.064046}, D94, 064046

\bibitem[\protect\citeauthoryear{Liu, Guo, Cai  \& Kim}{Liu
  et~al.}{2020}]{Liu:2020cds}
Liu L.,  Guo Z.-K.,  Cai R.-G.,   Kim S.~P.,  2020, \mn@doi [Phys. Rev. D]
  {10.1103/PhysRevD.102.043508}, 102, 043508

\bibitem[\protect\citeauthoryear{Macedo}{Macedo}{2018}]{Macedo:2018txb}
Macedo C.~F.,  2018, \mn@doi [Phys. Rev. D] {10.1103/PhysRevD.98.084054}, 98,
  084054

\bibitem[\protect\citeauthoryear{{Max} \& {Perkins}}{{Max} \&
  {Perkins}}{1971}]{1971PhRvL..27.1342M}
{Max} C.,  {Perkins} F.,  1971, \mn@doi [Physical Review Letters]
  {10.1103/PhysRevLett.27.1342}, \href
  {http://adsabs.harvard.edu/abs/1971PhRvL..27.1342M} {27, 1342}

\bibitem[\protect\citeauthoryear{Molina, Pani, Cardoso  \& Gualtieri}{Molina
  et~al.}{2010}]{Molina:2010fb}
Molina C.,  Pani P.,  Cardoso V.,   Gualtieri L.,  2010, \mn@doi [Phys. Rev.]
  {10.1103/PhysRevD.81.124021}, D81, 124021

\bibitem[\protect\citeauthoryear{Okounkova}{Okounkova}{2020}]{Okounkova:2020rqw}
Okounkova M.,  2020, \mn@doi [Phys. Rev. D] {10.1103/PhysRevD.102.084046}, 102,
  084046

\bibitem[\protect\citeauthoryear{Okounkova, Stein, Scheel  \&
  Hemberger}{Okounkova et~al.}{2017}]{Okounkova:2017yby}
Okounkova M.,  Stein L.~C.,  Scheel M.~A.,   Hemberger D.~A.,  2017, \mn@doi
  [Phys. Rev.] {10.1103/PhysRevD.96.044020}, D96, 044020

\bibitem[\protect\citeauthoryear{Pani}{Pani}{2013}]{Pani:2013pma}
Pani P.,  2013, \mn@doi [Int. J. Mod. Phys.] {10.1142/S0217751X13400186}, A28,
  1340018

\bibitem[\protect\citeauthoryear{Pani \& Loeb}{Pani \&
  Loeb}{2013}]{Pani:2013hpa}
Pani P.,  Loeb A.,  2013, \mn@doi [Phys.Rev.] {10.1103/PhysRevD.88.041301},
  D88, 041301

\bibitem[\protect\citeauthoryear{Pani, Cardoso, Gualtieri, Berti  \&
  Ishibashi}{Pani et~al.}{2012a}]{Pani:2012vp}
Pani P.,  Cardoso V.,  Gualtieri L.,  Berti E.,   Ishibashi A.,  2012a, \mn@doi
  [Phys. Rev. Lett.] {10.1103/PhysRevLett.109.131102}, 109, 131102

\bibitem[\protect\citeauthoryear{Pani, Cardoso, Gualtieri, Berti  \&
  Ishibashi}{Pani et~al.}{2012b}]{Pani:2012bp}
Pani P.,  Cardoso V.,  Gualtieri L.,  Berti E.,   Ishibashi A.,  2012b, \mn@doi
  [Phys.Rev.] {10.1103/PhysRevD.86.104017}, D86, 104017

\bibitem[\protect\citeauthoryear{Peters}{Peters}{1964}]{Peters:1964zz}
Peters P.,  1964, \mn@doi [Phys. Rev.] {10.1103/PhysRev.136.B1224}, 136, B1224

\bibitem[\protect\citeauthoryear{Plascencia \& Urbano}{Plascencia \&
  Urbano}{2018}]{Plascencia:2017kca}
Plascencia A.~D.,  Urbano A.,  2018, \mn@doi [JCAP]
  {10.1088/1475-7516/2018/04/059}, 1804, 059

\bibitem[\protect\citeauthoryear{Price}{Price}{1972}]{Price:1971fb}
Price R.~H.,  1972, \mn@doi [Phys. Rev.] {10.1103/PhysRevD.5.2419}, D5, 2419

\bibitem[\protect\citeauthoryear{Ramazano\u~glu \& Pretorius}{Ramazano\u~glu \&
  Pretorius}{2016}]{Ramazanoglu:2016kul}
Ramazano\u~glu F.~M.,  Pretorius F.,  2016, \mn@doi [Phys. Rev. D]
  {10.1103/PhysRevD.93.064005}, 93, 064005

\bibitem[\protect\citeauthoryear{Regge \& Wheeler}{Regge \&
  Wheeler}{1957}]{Regge:1957td}
Regge T.,  Wheeler J.~A.,  1957, \mn@doi [Phys. Rev.]
  {10.1103/PhysRev.108.1063}, 108, 1063

\bibitem[\protect\citeauthoryear{{Ruderman} \& {Sutherland}}{{Ruderman} \&
  {Sutherland}}{1975}]{1975ApJ...196...51R}
{Ruderman} M.~A.,  {Sutherland} P.~G.,  1975, \mn@doi [\apj] {10.1086/153393},
  \href {http://adsabs.harvard.edu/abs/1975ApJ...196...51R} {196, 51}

\bibitem[\protect\citeauthoryear{Siemonsen \& East}{Siemonsen \&
  East}{2020}]{Siemonsen:2019ebd}
Siemonsen N.,  East W.~E.,  2020, \mn@doi [Phys. Rev.]
  {10.1103/PhysRevD.101.024019}, D101, 024019

\bibitem[\protect\citeauthoryear{Will}{Will}{2014}]{Will:2014kxa}
Will C.~M.,  2014, \mn@doi [Living Rev. Rel.] {10.12942/lrr-2014-4}, 17, 4

\bibitem[\protect\citeauthoryear{Witek, Cardoso, Ishibashi  \& Sperhake}{Witek
  et~al.}{2013}]{Witek:2012tr}
Witek H.,  Cardoso V.,  Ishibashi A.,   Sperhake U.,  2013, \mn@doi [Phys.Rev.]
  {10.1103/PhysRevD.87.043513}, D87, 043513

\bibitem[\protect\citeauthoryear{Witek, Gualtieri, Pani  \& Sotiriou}{Witek
  et~al.}{2019}]{Witek:2018dmd}
Witek H.,  Gualtieri L.,  Pani P.,   Sotiriou T.~P.,  2019, \mn@doi [Phys. Rev.
  D] {10.1103/PhysRevD.99.064035}, 99, 064035

\bibitem[\protect\citeauthoryear{Yagi \& Stein}{Yagi \&
  Stein}{2016}]{Yagi:2016jml}
Yagi K.,  Stein L.~C.,  2016, \mn@doi [Class. Quant. Grav.]
  {10.1088/0264-9381/33/5/054001}, 33, 054001

\bibitem[\protect\citeauthoryear{Yunes, Pani  \& Cardoso}{Yunes
  et~al.}{2012}]{Yunes:2011aa}
Yunes N.,  Pani P.,   Cardoso V.,  2012, \mn@doi [Phys. Rev. D]
  {10.1103/PhysRevD.85.102003}, 85, 102003

\bibitem[\protect\citeauthoryear{Zaja\v~cek et~al.,}{Zaja\v~cek
  et~al.}{2019}]{Zajacek:2018vsj}
Zaja\v~cek M.,  et~al., 2019, \mn@doi [J.\ Phys.\ Conf.\ Ser.]
  {10.1088/1742-6596/1258/1/012031}, 1258, 012031

\bibitem[\protect\citeauthoryear{Zel'dovich}{Zel'dovich}{1971}]{zeldovich1}
Zel'dovich Y.~B.,  1971, Pis'ma Zh. Eksp. Teor. Fiz., 14, 270 [JETP Lett.
  {\bf14}, 180 (1971)]

\bibitem[\protect\citeauthoryear{Zel'dovich}{Zel'dovich}{1972}]{zeldovich2}
Zel'dovich Y.~B.,  1972, Zh. Eksp. Teor. Fiz, 62, 2076 [Sov.Phys. JETP {\bf
  35}, 1085 (1972)]

\bibitem[\protect\citeauthoryear{Zerilli}{Zerilli}{1970}]{Zerilli:1971wd}
Zerilli F.,  1970, \mn@doi [Phys. Rev. D] {10.1103/PhysRevD.2.2141}, 2, 2141

\bibitem[\protect\citeauthoryear{Zerilli}{Zerilli}{1974}]{Zerilli:1974ai}
Zerilli F.~J.,  1974, \mn@doi [Phys. Rev.] {10.1103/PhysRevD.9.860}, D9, 860

\bibitem[\protect\citeauthoryear{Zhu \& Osburn}{Zhu \&
  Osburn}{2018}]{Zhu:2018tzi}
Zhu R.,  Osburn T.,  2018, \mn@doi [Phys. Rev.] {10.1103/PhysRevD.97.104058},
  D97, 104058

\bibitem[\protect\citeauthoryear{Zilh\~ao, Cardoso, Herdeiro, Lehner  \&
  Sperhake}{Zilh\~ao et~al.}{2012}]{Zilhao:2012gp}
Zilh\~ao M.,  Cardoso V.,  Herdeiro C.,  Lehner L.,   Sperhake U.,  2012,
  \mn@doi [Phys. Rev.] {10.1103/PhysRevD.85.124062}, D85, 124062

\bibitem[\protect\citeauthoryear{Zilh\~ao, Cardoso, Herdeiro, Lehner  \&
  Sperhake}{Zilh\~ao et~al.}{2014}]{Zilhao:2013nda}
Zilh\~ao M.,  Cardoso V.,  Herdeiro C.,  Lehner L.,   Sperhake U.,  2014,
  \mn@doi [Phys. Rev.] {10.1103/PhysRevD.89.044008}, D89, 044008

\bibitem[\protect\citeauthoryear{Macedo}{rdw}{}]{rdweb}


\makeatother
\end{thebibliography}

\appendix

\section{Ringdown from the collision of charged BHs}\label{app:ringdown}

We can gain some insight into the ringdown stage of the collision of two electrically charged BH by looking at the 
simplified problem of a charged particle plunging into a charged BH. This setup was explored in previous 
work~\citep{Zerilli:1974ai,Johnston:1973cd,Cardoso:2003cn,Cardoso:2016olt}. Here we outline the main aspects to 
illustrate that --~neglecting EM-plasma interactions~-- the ringdown stage 
is contaminated by other modes, namely the EM one.

Consider a charged particle falling radially into a charged BH. Due to the symmetry of the problem, the particle only excites the polar sector of the perturbations. The metric, therefore, can be written as
\begin{equation}
g_{ab}=g_{ab}^{(0)}+h_{ab},
\end{equation}
with $g_{ab}^{(0)}={\rm diag}\{-f,f^{-1},r^2,r^2\sin^2\theta\}$, with $f=1-2M/r+Q^2/r^2$, where $M$ is the mass and $Q$ 
the BH charge. The metric perturbation $h_{ab}$ induced by the falling particle can be decomposed into spherical 
harmonics. We can study the perturbation in the Regge-Wheeler gauge~\citep{Regge:1957td}, 
\begin{equation}
h_{ab}=
\left(
\begin{array}{c c c c}
e^{v}H_0& H_1         & 0     & 0\\
H_1     & e^{-v}{H_2} & 0     & 0\\
0       & 0           & r^2 K & 0\\
0       & 0           & 0     & r^2\sin^2\theta K
\end{array}
\right)Y_{lm} (\theta,\phi),
\label{eq:metrip}
\end{equation}
with $Y_{lm} (\theta,\phi)$ being the standard spherical harmonics. The electromagnetic perturbations can be also 
expanded in harmonics, being described by the perturbation of the vector potential $\delta A_a =(\delta A_0,\delta 
A_1,0,0)Y_{lm}$.
Finally, one can also decompose the particle stress-energy tensor into tensorial 
harmonics as~\citep{Zerilli:1971wd}
\begin{equation}
	T_{ab}=\left(
	\begin{array}{c c c c}
	A_{0} & \frac{i}{\sqrt{2}}A_{(1)} & 0 & 0\\
	\frac{i}{\sqrt{2}}A_{(1)} & A & 0 & 0\\
	0 & 0 & 0 & 0\\
	0 & 0 & 0 & 0
	\end{array}
	\right)Y_{lm},
\end{equation}
where the radial functions $A_{0}$, $A_{(1)}$, and $A$ depend on the particle's motion. 
By plugging this into Einstein-Maxwell equation and upon linearization, we can find equations describing the 
perturbations of the metric perturbations and of the vector potential. We direct the reader to 
Ref.~\citep{Cardoso:2016olt} for more details.

\begin{figure*}
	\includegraphics[width=\linewidth]{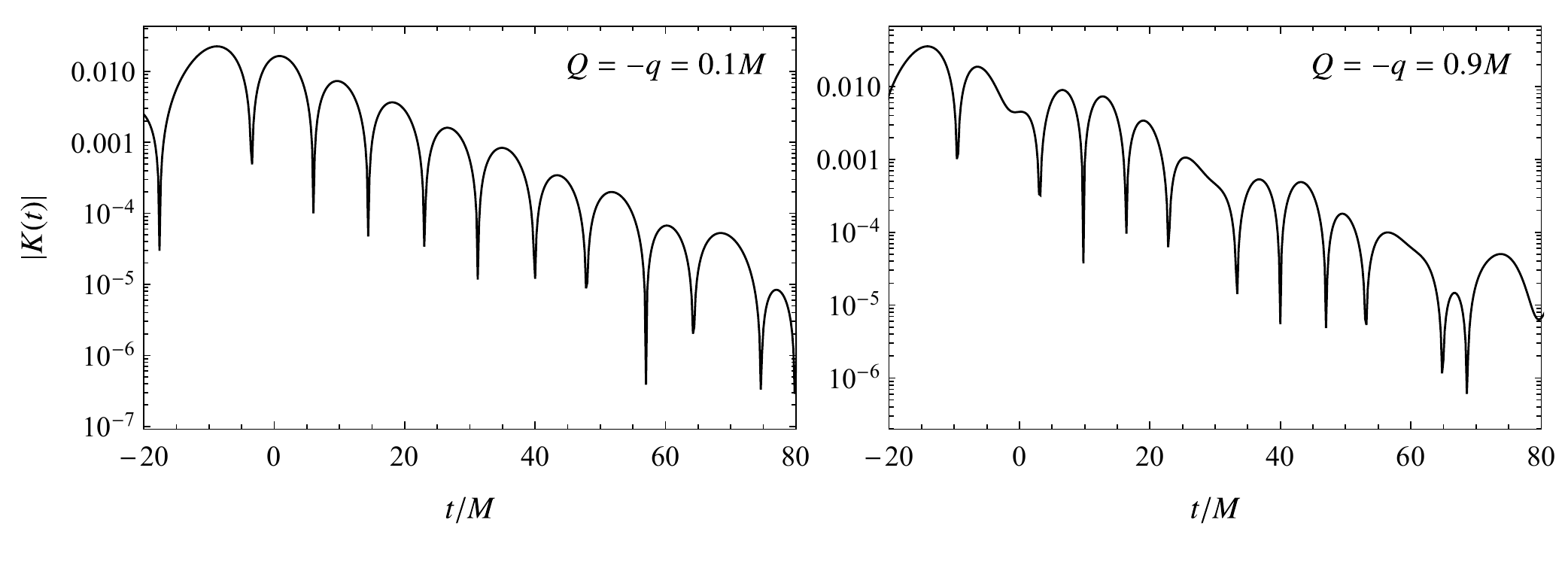}
	\caption{Ringdown produced by the collision of charged BH with a charged particle (taking only the dominant 
quadrupole term). The left and right panels correspond to the case of 
relatively small and large charges, respectively. In the former case the signal is essentially dominated by the 
gravitational-led mode, whereas for highly charged configurations the signal is contaminated by the additional EM-led 
mode.}
	\label{fig:ring}
\end{figure*}

\begin{figure}
	\includegraphics[width=\linewidth]{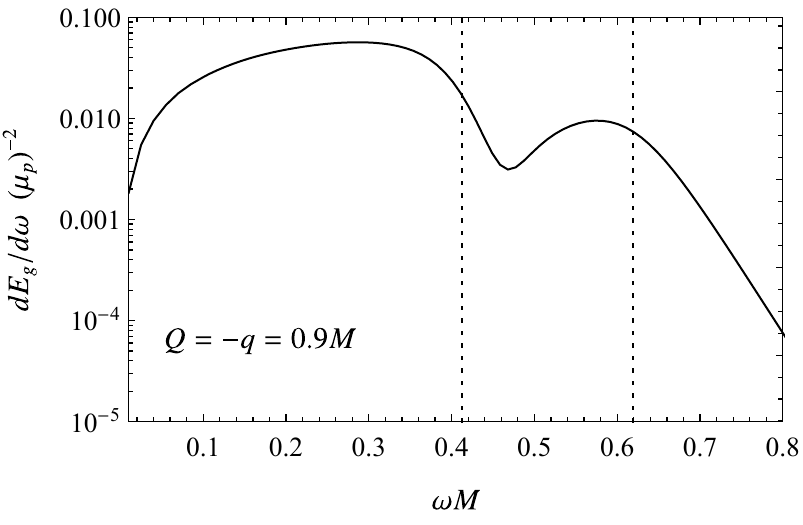}
	\caption{Gravitational energy flux as a function of the frequency $\omega$ for the plunge of a particle 
with mass $\mu_p$ and charge $q$ into a BH with charge $Q$ (taking only the dominant quadrupole term). The vertical 
dotted lines mark the real part of the gravitational and EM fundamental modes of the corresponding Reissner-Nordstrom 
BH.}
	\label{fig:flux}
\end{figure}

By considering plunging processes, we can investigate the waveform emitted during the collision. The waveform presents 
the characteristic ringdown in which the signal oscillates according to the quasinormal mode frequency of the BH 
decaying exponentially in time. In Fig.~\ref{fig:ring} we show the GWs emitted\footnote{Note that here we are using 
the $K$ function of 
Eq.~\eqref{eq:metrip} to represent the 
GW. We remark that $K$ is related to the (gauge-invariant) Zerilli function asymptotically by a derivative.} from the 
collision of a BH with 
charge $Q$ with a charged particle with charge $q=-Q$. 
The left panel shows the case of relatively small charges ($Q=-q=0.1M$), in which the 
ringdown contains basically the gravitational-led fundamental quasinormal mode of the central Reissner-Nordstrom BH. 
For higher charges (right panel of Fig.~\ref{fig:ring}, in which $Q=-q=0.9M$), the signal is less regular, being 
contaminated by additional modes. The waveform may be thought as the
superposition of two types of modes: the standard gravitational-led one and the EM-led one, which is sufficiently excited in the 
high-charge (i.e. large-coupling) case and it therefore appears in the GW signal. In fact, by considering an expansion of the form
\begin{equation}
	K(t)\approx \sum_n K_n e^{-i\omega_n t},
\end{equation}
one can use the Prony method to find that the expansion is predominantly given by the frequencies of the fundamental gravitational-led (with frequency $\omega M=0.414-0.088i$) and EM ($\omega M=0.620-0.096i$) modes.
The superposition of modes can also be seen by analyzing the 
power spectrum of the GW flux, as shown in Fig.~\ref{fig:flux}, where the vertical lines mark the real parts 
gravitational-led and the EM-led modes.

The latter case is what one may expect when the BH is isolated, without the presence of plasma. As discussed in the 
main text, the main effect of the EM-plasma coupling is to provide an effective mass term for the EM mode, 
suppressing its excitation in the GW signal.

\bsp	
\label{lastpage}

\end{document}